# Site-Selective Yttrium Substitution in $Ti_3AlC_2$ Enables Interlayer Engineering and $Li^+$ Transport in $Ti_3C_2T_x$ cathodes for High-Power Energy Storage

Tetiana Boichuk[1*], Andrii Boichuk[1,2*], Mahesh Eledath Changarath[1], João Fonseca[1], Said Agouram[3], Marie Finas[1], Alejandro Molina-Sánchez[1] and Juan F. Sánchez-Royo[1†]

[1] ICMUV, Instituto de Ciencia de Materiales, Universidad de Valencia, 46071 Valencia, Spain

[2] King Danylo University, 76000, Ivano-Frankivsk, Ukraine

[3] Department of Applied Physics and Electromagnetism, University of Valencia, 46100 Valencia

[*]Corresponding author: Andrii.Boichuk@uv.es (Andrii Boichuk)

[*]Corresponding author: Tetiana.Boichuk@uv.es (Tetiana Boichuk)

[†]Corresponding author: Juan.F.Sanchez@uv.es (Juan F. Sánchez-Royo)

## Abstract

Engineering rapid ion transport within layered MXenes through doping is an effective way to improve ion transport in electrodes for batteries. However, most reported methods modify Ti-C MXenes during or after etching and do not change the crystal structure of the parent MAX phase. In this work, we introduce a precursor-level strategy by substitutionally doping the $Ti_3AlC_2$ MAX precursor with a small amount of Yttrium during top-down MAX preparation. This allows permanent structural changes that are retained in the resulting $Ti_3C_2T_x$ MXene. Density functional theory (DFT) calculations identify the outer Ti(4f) site, facing the Al layer, as the preferred incorporation site for Y under the Al-excess synthesis conditions used here, rather than the inner Ti(2a) or Al(2b) sites. This site-specific substitution is consistent with the experimentally observed lattice expansion and carbide-like Y bonding and provides a microscopic basis for the structural changes retained in the resulting $Ti_3C_2T_x$ MXene. Y incorporation expands the MXene interlayer spacing due to an increase in the amount of hydroxyl surface terminations, creating more favorable pathways for lithium-ion transport. Comprehensive structural characterization (including XRD, XPS and TEM) confirms these modifications. As a result, the Y-doped $Ti_3C_2T_x$ cathode shows diffusion-controlled charge storage with a $Li^+$ diffusion coefficient of $10^{-9}$-$10^{-11}$ $cm^2/s$. It delivers a reversible capacity of about 130 mAh/g at 0.2 C and retains 74 mAh/g after 1000 cycles at 2 C with nearly 100%

Coulombic efficiency. The electrode also achieves a high power density of up to 3970 W/kg, outperforming the undoped MXene and previously reported multilayered MXene cathodes. These results show that substitutional doping of the MAX precursor is a promising strategy for controlling the interlayer structure of MXenes and improving ion transport, providing a new route for the development of high-power energy-storage electrodes.

## Introduction

The fast development of portable electronics, electric vehicles, and renewable energy systems has increased the demand for advanced energy storage devices, in particular batteries and supercapacitors. Modern applications require electrode materials that can deliver both high power density and fast charge–discharge capability of energy systems. Therefore, designing electrode materials with rapid ion transport and high charge storage capacity is very important, regardless of the type of energy storage - Li, Na, or others. Layered materials have attracted strong interest for energy storage because of their structure, large surface area, short ion diffusion paths, and compatibility with different ions, practically independent of their size. As we have shown in our previous works [1-3], layered 2D transition-metal oxides exhibit improved electrochemical performance compared to bulk materials, with optimized pseudocapacitance, providing high capacity and cycling stability. However, the semiconducting nature of these materials limits ion kinetic during electrochemical reactions [4-6] and requires adding of conductive additives, which decrease specific performance.

MXenes, especially $Ti_3C_2T_x$, are considered promising electrode materials due to their metallic conductivity, layered structure, possibilities of surface functionalization, and morphology tuning [7-9]. Their high electrical conductivity allows fast electron transport, which helps reduce internal resistance and improves ion diffusion [10-12]. However, high conductive electrodes based on single flakes of $Ti_3C_2T_x$, as shown in [13], prepared using traditional fabrication methods, cause restacking of 2D flakes, which blocks ion transport and reduces performance in thick films. This fact is probably behind the increasing interest on multilayered MXene configuration for applications in energy storage with high power density [14-16]. However, remaining problems, such as low electrical conductivity and stability of interlayer spacing during long-term electrochemical reactions, should be solved. Doping is an effective way for increasing the interlayer spacing is an effective strategy to improve ion transport and electrolyte access. Review [17] presents the study of heteroatom doping (N, S, P) of Mxenes for a wide range of applications, showing that heteroatom doping enhanced the interlayer spacing of MXene layers, which further improved the charge storage ability. Other works [18-21] describe a positive influence of doping on the electronic, electrochemical properties, and stability of Ti-based and other Mxenes. In the context of energy storage, Ru-doped MXene [22] shows improved electrochemical performance compared to pure Ti-C MXenes, showing that interlayer engineering can reduce diffusion resistance and enhance electrochemical performance. However, the examples of doping described above were performed with already exfoliated MXenes through the addition of dopants during etching and post-etching procedures, changing the functionalization of the surface without replacing the M element of the MAX phase. At the same time, doping of Ti Mxene using a low amount of heterovalent element with a bigger atomic size (like Yttrium) during MAX synthesis may substitute Ti atoms with further effects: forming of additional microstrains (influence on interlayer distance after etching) as well as changes in the electronic structure (increasing of electrical conductivity). Such dopant-induced modifications provide a direct pathway to tune the MXene interlayer environment, leading to a higher ion diffusion coefficient. In this work, we investigate the influence of low-concentration Yttrium doping within the parent MAX phase on the resulting structure, interlayer spacing, and electrochemical performance of T-Y-C MXene cathodes for lithium batteries. This study is supported by density functional theory (DFT) calculations, aimed to identify the preferred atomic positions of Y, which are then correlated with comprehensive structural characterization and electrochemical transport kinetic.

## Experimental section

### *Synthesis and processing of Y-doped MAX and Mxenes*

The Y-doped Ti-Al-C MAX phase has been synthesized using high-purity metallic powders of Ti, Y, Al, and commercial graphite in a molar ratio 2,7:0,3:1,1:2,0. Firstly, a mix of powders has been added to a 50 ml stainless steel jar and milled using a planetary ball mill (600 rpm, 5 hours). After, the resulting precursor has been placed into a tube furnace and calcined in Ar atmosphere (heating rate 3 degrees per minute, calcination temperature 1420, time - 2 hours ). Post-processing and etching of the MAX phase have been realized using HF/HCl according to the protocol [23]. After washing the etched solution, the multilayered MXene (powder) (**Fig.1a**) was collected using vacuum filtration, dried in a vacuum oven, and stored in a glovebox until electrode preparation. For the comparison, the undoped $Ti_3AlC_2$ MAX phase has been prepared using the same precursors and synthesis conditions.

### *Computational methods*

Spin-polarized DFT calculations used VASP 6.5.0 [24] with the PBE functional [25] and projector augmented-wave (PAW) potentials [9, 10] (Ti pv, Y sv, Al, C), a 520 eV cutoff, and Γ-centered 3×3×1 meshes for the 108-atom Ti3AlC2 supercell. Structures were relaxed to forces below 0.01 eV/Å and evaluated in static calculations. The solution energy of Y followed Nie et al. [26] and Burr et al. [27], with hexagonal close-packed (hcp) Ti, face-centered cubic (fcc) Al, and hcp Y as reservoirs. The site ranking across chemical environments followed Wang et al. [28], with 30 competing Ti–Al–Y–C phases computed at the same settings. Two-Y arrangements were enumerated with enumlib [29] and pre-ranked with a machine-learned interatomic potential before DFT (SI). Random (Ti,Y) solid solutions were modeled with special quasi-random structures (SQS) [30] in the same supercell at y = 0.056–0.222. XRD patterns were simulated for Cu Kα with pymatgen [31].

*Structure and morphology characterization*

X-ray diffraction (XRD) patterns were recorded using a Bruker D8 ADVANCE A25 diffractometer with Cu-Kα radiation (λ = 1.54 Å) in the 2θ range of 5–80° and a step size of 0.02°. The surface morphology of the samples was examined by field emission scanning electron microscopy (SCIOS 2, 10 kV) with the Oxford Instruments EDS module. High-resolution transmission electron microscopy (HR TEM) and EDX measurements were conducted using a TECNAI G2 F20 microscope operating at 200 kV. X-ray photoelectron spectroscopy (XPS) measurements were performed using a Thermo Scientific K-Alpha system with monochromatic Al-Kα radiation (1486.6 eV, base pressure ~$1\times10^{-10}$ mbar). Raman spectroscopy measurements were carried out using a Horiba Scientific Xplora µ-Raman spectrometer equipped with a 532 nm excitation laser. The laser beam was focused onto the sample, producing an illuminated area of approximately 1 $mm^2$. Spectra were acquired at a laser power of 100 mW with an acquisition time of 5 s and two accumulations per measurement. All measurements were performed under ambient conditions. To assess the homogeneity and reproducibility of the samples, 5–7 spectra were collected from different locations on each specimen.

*Electrochemical characterization*

Electrochemical characterization, including charge–discharge cycling and the Galvanostatic Intermittent Titration Technique (GITT), was performed on CR2032 coin cells using an ARBIN LBT21084 battery test system at room temperature within a voltage window of 0–3 V. The cathode was prepared by mixing multilayered Y-doped MXene powder (80 wt%), carbon black (15 wt%), and polyvinylidenefluoride (PVDF, 5 wt%). The resulting electrode had a diameter of 11 mm and an active material loading of 3.9 mg/cm$^2$. Metallic lithium was used as the anode, and a $LiPF_6$ solution in EC/DMC served as the electrolyte. Cyclic voltammetry (CV) measurements were carried out using a Gamry Reference 5000E potentiostat at scan rates ranging from 2 to 200 mV/s. Electrochemical impedance spectroscopy (EIS) was performed using the same instrument during discharge of the coin cell with a voltage step of 0.5 V, starting from 3 V down to full discharge (0 V). The impedance spectra were recorded at fixed voltages over a frequency range of 20 kHz to 0.01 Hz using 63 frequency points, with an AC voltage amplitude of 5 mV. For the GITT measurements, the cell was subjected to a series of 300 s current pulses, each followed by a 900 s relaxation period under open-circuit conditions, during which the cell potential was monitored.

## RESULTS AND DISCUSSION

**Fig. 1b** presents the XRD patterns of the Y-doped $Ti_3AlC_2$ MAX precursor and the corresponding $Ti_3C_2T_x$ MXene after chemical etching. The diffraction pattern of the as-synthesized precursor is dominated by the characteristic reflections of the $Ti_3AlC_2$ MAX phase, confirming its successful formation. The most intense peaks at approximately 9.6°, 19°, 39°, 48°, 56°, 69°, and 75° are indexed to the (002), (004), (104), (107), (109), (201), and (204) planes of $Ti_3AlC_2$, respectively. Besides the main MAX phase, weak reflections at approximately 42° and 61° correspond to the (200) and (220) planes of TiC, indicating the presence of a minor TiC impurity formed during high-temperature synthesis [32,33]. In addition, a weak diffraction peak at approximately 29–30°, together with low-intensity reflections near 53°, 57–58°, 69°, and 76°, can be tentatively indexed to the (222), (611), (622), (800), and (840) planes of cubic $Y_2O_3$ [34-36], which was formed from unreacted metallic Y after calcination in air environment. The formation of this oxide phase suggests that a small fraction of yttrium remained unincorporated into the MAX lattice during sintering, so the real amount of Y in the MAX phase will be confirmed by EDX. Compared to the undoped Ti-C MAX sample **(Fig. SI 1)**, based on XRD, we can notice that, regarding the phase content, the only difference is the presence of a small amount of $Y_2O_3$. A magnified view of the (104) reflection is presented in **Fig. SI 2**. Compared with undoped MAX phase, the Y-containing sample exhibits a clear shift of the (104) peak toward lower 2θ values together with slight peak broadening, which corresponds to an increase in the interplanar spacing, indicating lattice expansion after Y addition.

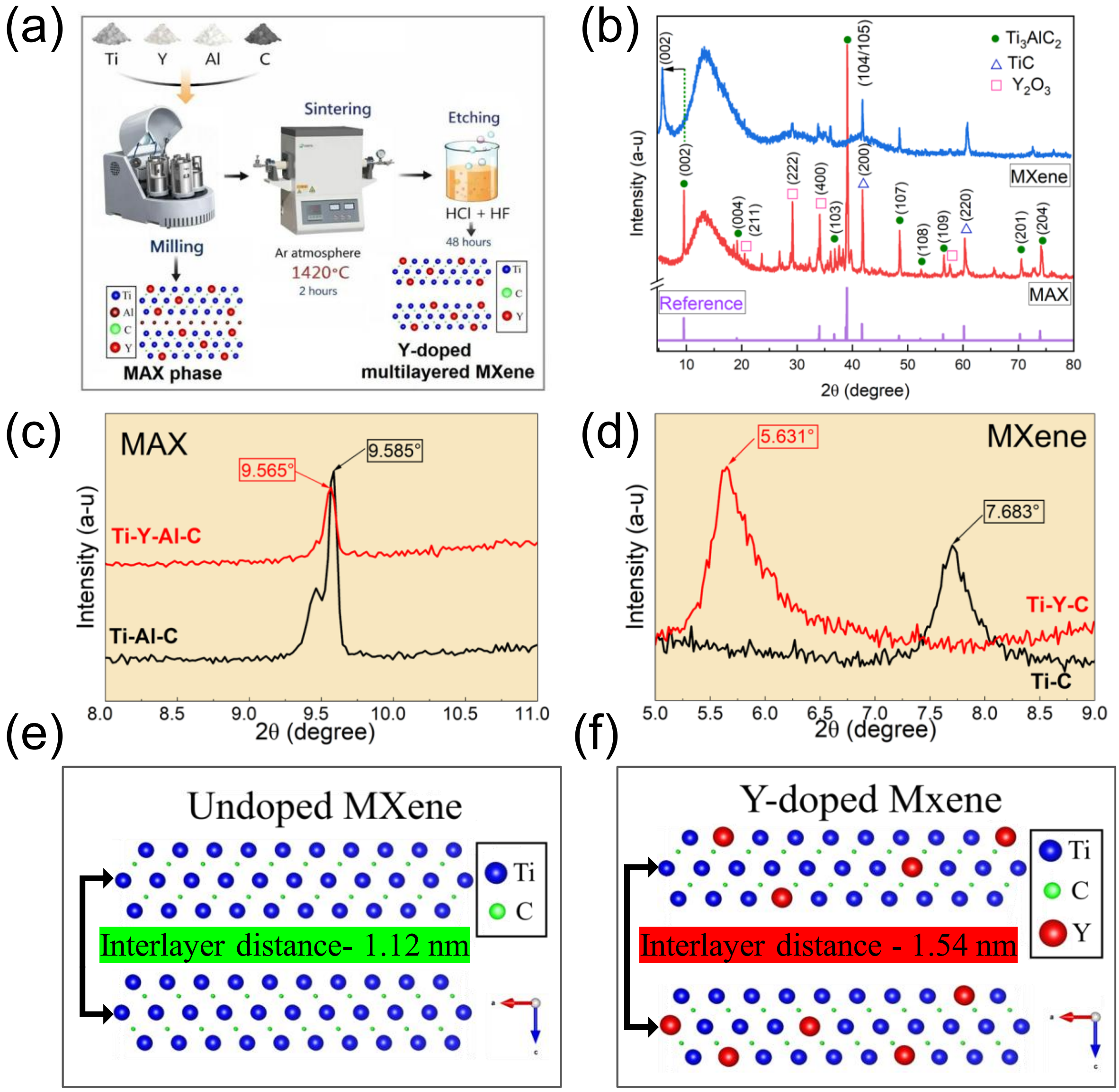

**Figure 1.** Synthesis and structural evolution of Y-doped $Ti_3C_2T_x$ MXene. (a) Scheme of Y-doped MAX synthesis and further processing. (b) XRD patterns of Y-doped $Ti_3C_2T_x$ MAX phase before and after HF/HCl etching. (c) Comparison of the (002) reflection of undoped and Y-doped MAX phases. (d) Comparison of the (002) reflection of undoped and Y-doped $Ti_3C_2T_x$ MXenes after etching. Comparison of interlayer spacing for undoped (e) and Y-doped (f) $Ti_3C_2T_x$ MXenes calculated from the (002) diffraction peak.

Because the (104) reflection is sensitive to the geometry of the Ti–C framework, the observed behavior is consistent with the incorporation of the larger Y atoms into Ti positions After etching of Y-doped MAX phase **(Fig.1b)**, besides to (002) shift, the intensity of the characteristic $Ti_3AlC_2$ (104) reflection at approximately 39° decreases drastically, demonstrating the effective elimination of the parent MAX phase. The persistence of weak TiC reflections confirms that TiC remains chemically stable during the etching process. Likewise, the weak reflection near 29,2°, attributed to cubic $Y_2O_3$, remains essentially unchanged after etching, indicating that this oxide phase is inert toward HF and survives the selective removal of Al from the MAX structure. Although weak TiC and $Y_2O_3$ reflections are still visible in the XRD pattern of the etched sample, this is expected since the analysis was performed after the initial high-speed centrifugation, which can retain fine impurity particles in the MXene fraction. To further improve the purity, the suspension was subjected to low-speed centrifugation, allowing the larger, denser TiC and $Y_2O_3$ particles to sediment while the multilayered Y-containing MXene mostly remained dispersed in the supernatant. This additional purification step yielded a MXene suspension with a substantially lower impurity content. The position of the (002) peak for Y-doped Ti-C MAX phase is shifted to lower angles, indicating the incorporation of Y with a larger atomic size into the lattice of MAX, with the corresponding c-lattice constant increasing **(Fig.1c)**. Following HCl/HF etching, the diffraction pattern changes significantly, confirming the successful conversion of the MAX phase into $Ti_3C_2T_x$ MXene. For undoped Ti-C MXene, as it is typically described in the literature [37,38], the characteristic (002) reflection shifts from approximately 9.58° to 7, 683°, accompanied by considerable peak broadening **(Fig.SI 3)**, indicating an increased interlayer spacing due to the removal of Al atomic layers and the introduction of surface terminations (-O, -OH, and -F). But, in the case of Y-doped compounds, we observe an untypical (002) shift (**Fig.1d**) - from 9,565° to 5,631°, which corresponds to a much bigger interlayer distance in Ti-Y-C multilayered MXenes **(Figs. 1e and 1f). Fig. SI 4** compares the same diffraction region after etching for the undoped and Y-doped MXene samples. In both cases, the characteristic (104) reflection of the MAX phase is no longer observed, confirming that the ordered MAX structure has been successfully transformed into $Ti_3C_2T_x$ MXene after selective Al removal. Assuming that Y substitutes Ti in the MAX lattice, the larger ionic radius of Y introduces local lattice distortion, which facilitates Al removal during etching and promotes the formation of surface terminations and interlayer hydration. Consequently, the $Ti_3C_2T_x$ layers exhibit greater expansion, which may boost ionic transport in the case of the use of multilayered Y-doped MXene as a cathode for batteries. into the MAX/MXene, additional structural/morphological studies have been conducted. SEM images of Ti-Y-Al-C MAX **(Fig.2a)** confirm the successful forming of a layered MAX structure with an average particle size of about 3-5 micrometers **(Fig.SI 5a)**, the uniform distribution of atoms in the MAX phase on the EDS image **(Fig.SI 3b)**, as well as the presence of $Y_2O_3$.

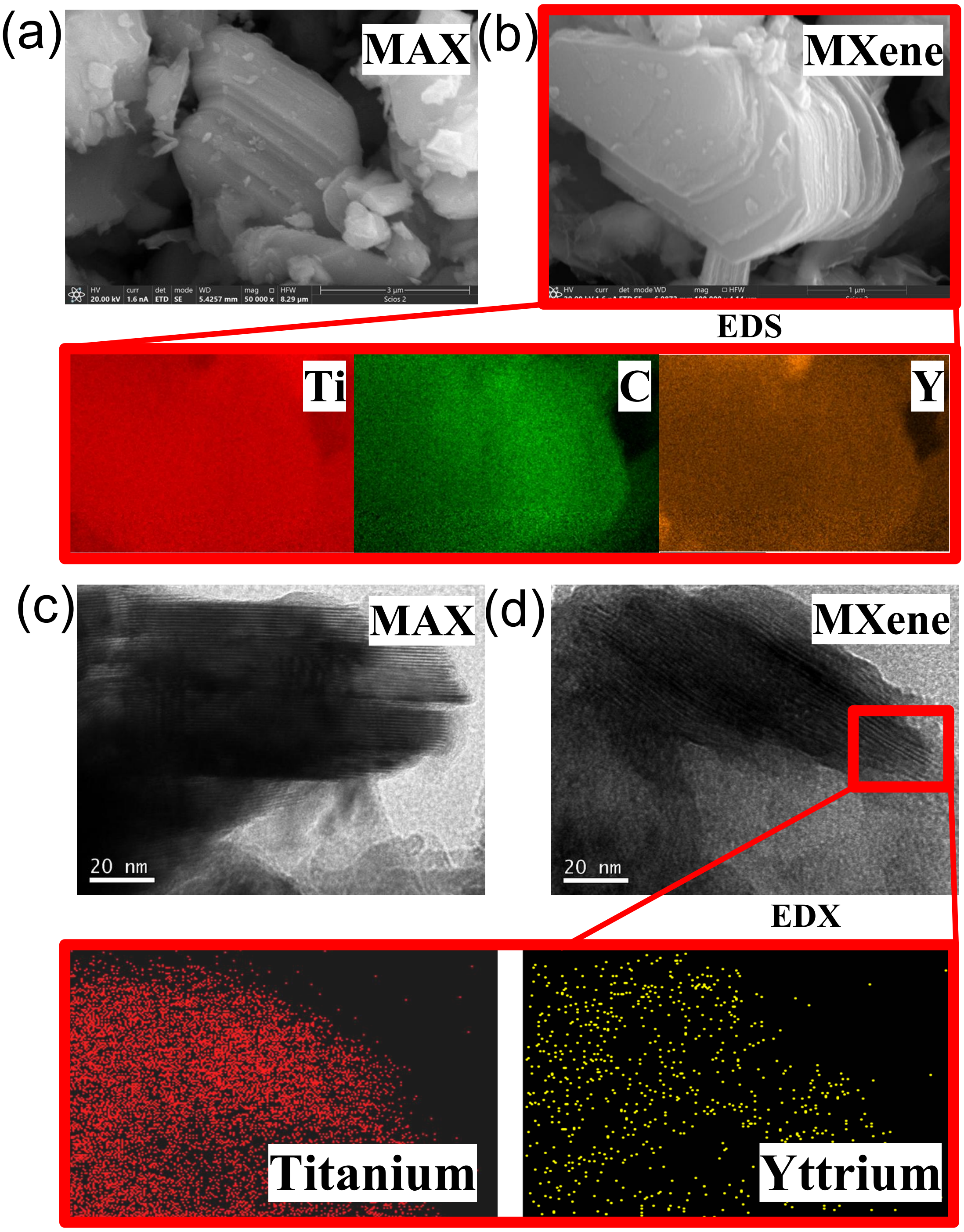

**Figure 2.** Morphological and structural characterization of Y-doped MAX and MXene. (a) SEM image and EDS elemental mapping of the Y-doped $Ti_3AlC_2$ MAX phase. (b) SEM image and elemental distribution of multilayered Y-doped $Ti_3AlC_2$ MXene after etching. (c) High-resolution TEM image of Y-doped $Ti_3AlC_2$ MAX phase showing the layered structure. (d) High-resolution TEM image of multilayered Y-doped $Ti_3AlC_2$ MXene together with EDX analysis confirming the presence of Y after etching.

SEM of multilayered Y-doped MXene **(Fig.2b)** exhibits accordion-like morphology with clearly visible Y atoms with uniform distribution in MXene flakes without a decrease of lateral dimensions after etching, compared to unetched MAX (**Fig.SI 6** shows the lateral MXene size about 5 micrometers).

In order to confirm the phenomena described above caused by the incorporation of Y Moreover, after delamination of multilayered MXene based on EDS of individual delaminated Y-doped MXene flakes (**Fig.SI 7**), we observe a uniform out-of-plane distribution of Yttrium. HR TEM of Y-doped MAX phase confirms XRD/SEM results with an interlayer distance of about 10.8 Å (**Fig.SI 8**). After etching, we observe typical interlayer distance increase due to the removal of Al (**Fig.2d**) with the presence of Y in multilayered MXene (EDX, **Fig.2d**). Quantitative EDX analysis of Y-doped MAX phase and multilayered Y - containing MXene (**Table SI 1**) shows that the Ti:Y ratio is stable (about 28) and does not depend on the presence of aluminum, functional groups, and etching procedures. Collectively, these results strongly support the successful incorporation of Y into the $Ti_3C_2T_x$ MXene, consistent with the proposed substitution of Ti by Y with a percentage of 3%. But, even this small amount of dopant significantly increases the interlayer distance of multilayered Y-doped Mxene compared to undoped MXene (**Fig.SI 9**), which is consistent with previously described XRD results.

To gain further insight into the influence of Y doping on the $Ti_3C_2T_x$ structure, XPS and Raman spectroscopy were carried out for both undoped and Y-doped MXene samples. These complementary techniques were employed to verify the chemical state of Y and to evaluate the structural modifications associated with its incorporation into the MXene lattice. **Fig. 3** (a-b) depicts the Ti 2p and C 1s core level XPS spectra acquired from Y-doped Mxene samples, respectively. The XPS spectra of the undoped Mxene sample are also shown in the figure for comparison. The spectra were deconvoluted by assuming a Voigt line shape and Shirley backgrounds. In the Ti 2p spectra, deconvolution reveals four distinct doublets that are attributed to Ti-C (454.9 eV), $Ti^{2+}$ (455.9 eV), $Ti^{3+}$ (457.6 eV), and $TiO_2$ (458.4 eV) [39, 41]. The C1s spectra exhibit four peaks, which are attributed to Ti-C-Ti (282 eV), Y-C-Ti (282.7 eV), C-C (284.8 eV), and C-O (286.3 eV) chemical bonds [39,40]. **Figs. 3d and 3e** presents Y 3d and F 1s core-level spectra, respectively. Deconvolution of the Y 3d spectrum reveals a single doublet, with the Y $3d_{5/2}$ peak located at 158.5 eV, which is attributed to the $C–Y–T_x$ bonding environment. The F 1s spectrum exhibits two characteristic peaks corresponding to $C–Ti–F_x$ (685.4 eV) and Al $(OF)_x$ (687.1 eV) surface terminations [39]. The weak AlFx signal is attributed to residual Al–F species, likely originating from incomplete removal of etching by-products during the washing process, as we did not observe the presence of Al in survey XPS spectra of Y-doped MXene after etching **(Fig.SI 10)**. Owing to its very low intensity, this component represents only a minor surface species compared to the predominant MXene surface terminations. The signal in the Cl 2p spectrum (**Fig.3c**) is attributed to residual chlorine species remaining from the HCl-based etching process.

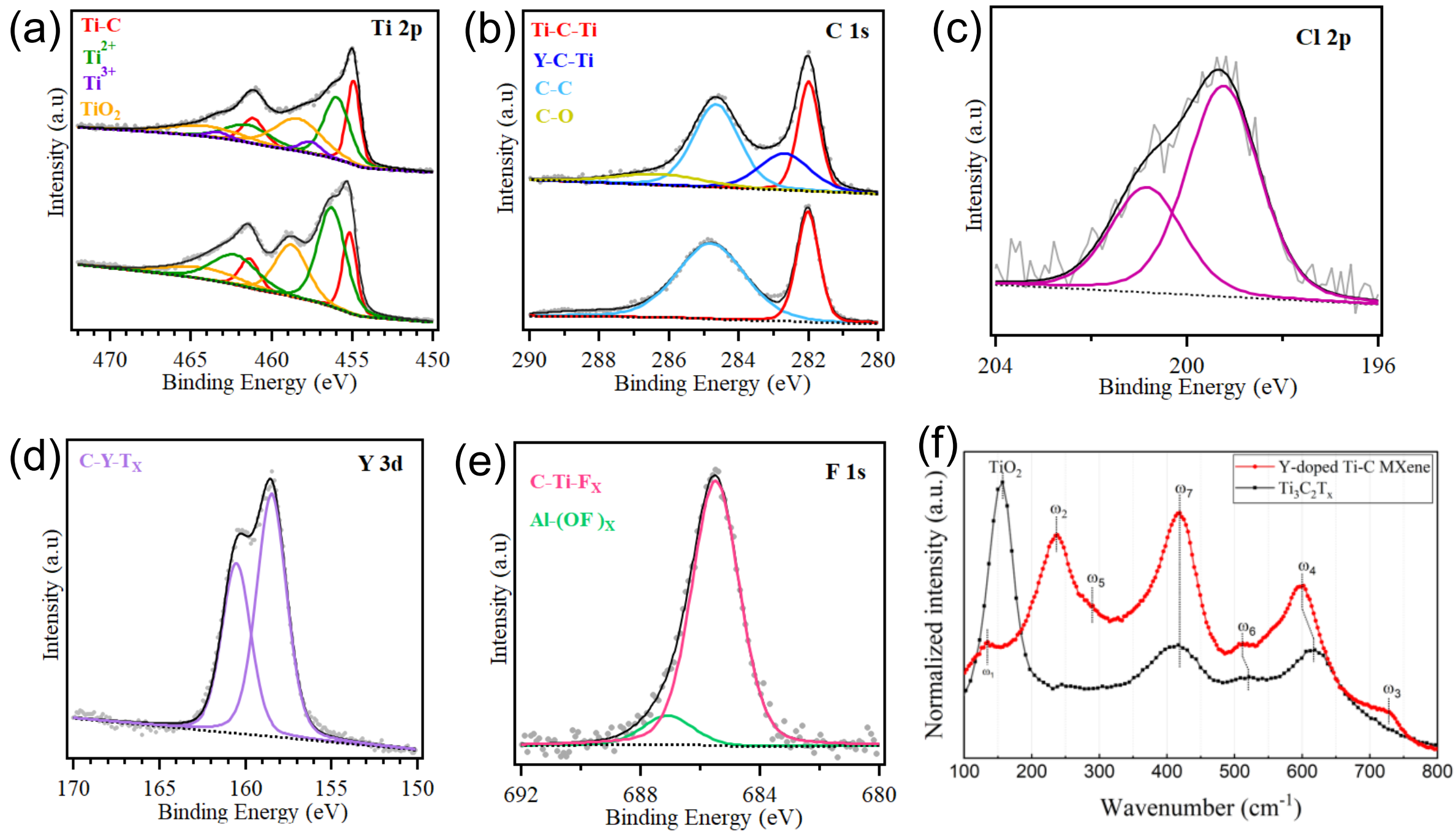

**Figure 3.** XPS and Raman characterization of Y-doped $Ti_3C_2T_x$ MXene. High-resolution XPS spectra of (a) Ti 2p, (b) C 1s, (c) Cl 2p, (d)Y 3d, and (e) F 1s regions for undoped and Y-doped $Ti_3C_2T_x$ MXenes. (f) Raman spectra of undoped and Y-doped $Ti_3C_2T_x$ MXenes showing the evolution of characteristic vibrational modes after Y incorporation.

The low-frequency Raman spectra of pristine and Y-doped $Ti_3C_2T_x$ MXenes reveal several characteristic changes associated with Y substitution into the lattice (**Fig.3f**). The spectrum of the pristine MXene is dominated by an intense in-plane $E_g$ mode at 155 $cm^{-1}$, assigned to $TiO_2$ vibrations [42], alongside bands at 418 $cm^{-1}$ ($\omega_7$, in-plane $E_g$, associated with -OH surface groups) and 620 $cm^{-1}$ ($\omega_4$, in-plane $E_g$, corresponding to carbon vibrations), as well as a weaker $A_{1g}$ band at 520 $cm^{-1}$ ($\omega_6$, out-of-plane vibration of OH groups) [43]. Upon Y doping, the $TiO_2$-related band at 155 $cm^{-1}$ disappears while a bump emerges at 135 $cm^{-1}$, attributed to in-plane $\omega_1$ ($E_g$) Ti/Y vibration. This suggests that Y substitution suppresses $TiO_2$ formation while enhancing the in-plane vibration of the skeleton. The most striking spectral evolution induced by Y substitution occurs in the 200–300 $cm^{-1}$ region, where an intense peak appears at 237 $cm^{-1}$ with a shoulder at 282 $cm^{-1}$. The main peak was assigned to the out-of-plane $\omega_2$ ($A_{1g}$) Ti/Y mode, and the shoulder to $\omega_5$ ($E_g$) in plan vibration of the OH mode [42,43]. Notably, the $\omega_2$ peak in the Y-doped samples is downshifted relative to the $\omega_2$ ($A_{1g}$) peak of the undoped MAX phase at 260 $cm^{-1}$ (**Fig.SI 11**), indicating that this vibrational mode in the MXene is directly impacted by Y substitution. The intense peak at 418 $cm^{-1}$, related to the in-plane $\omega_7$ ($E_g$) -OH mode, remains unchanged in position, indicating that this vibration is not strongly affected by Y incorporation. In contrast, the out-of-plane $A_{1g}$ -OH mode ($\omega_6$) downshifts from 520 to 510 $cm^{-1}$, while the in-plane $E_g$ mode ($\omega_4$) associated with carbon downshifts from 620 to 600 $cm^{-1}$. Both shifts are consistent with lattice perturbation induced by Y substitution, as Y is a heavier atom than Ti and forms softer bonds. Finally, a new bump emerges upon Y doping at 730 $cm^{-1}$, corresponding to out-of-plane $A_{1g}$ oxygen mode ($\omega_3$) [44], further supporting a modification of the surface groups. Compared with the literature, the weak intensity of the $\omega_1$, $\omega_6$, and $\omega_3$ components can be attributed to the non-resonant nature of the 532 nm excitation wavelength used in this study [45].

To provide a definitive microscopic justification for these experimental findings and unveil the exact atomic configuration of the Y-doped lattice, density-functional theory (DFT) calculations were performed. Thermodynamically, Y atoms could substitute either Ti or Al sites within the $Ti_3AlC_2$ framework **(Fig. 4a)**. We computed the solution energy of Y on each candidate site, following Nie et al. [26] and Burr et al. [27]: one Y atom from hcp Y replaces one Ti or Al atom on a host site, and the displaced atom returns to its elemental metal (hcp Ti or fcc Al). The solution energy is positive for every site and arrangement tested, from +1.17 to +2.93 eV per Y **(Table ST2)**, as expected from the 0.33 Å metallic radius mismatch between Ti and Y, which places this pair in the poorly mixing regime [46]. To evaluate thermodynamic stability, Fig. 4b shows the formation enthalpy of each doped cell relative to the convex hull of 30 competing Ti–Al–Y–C phases **(Table ST3)**. The Ti(4f) pair sits 13.5 meV/atom above the hull, while the Al(2b) pair lies 10.8 meV/atom above it. To test whether configurational entropy can stabilize the solid solution at different synthesis temperatures, we evaluated ideal mixing models. When pristine $Ti_3AlC_2$ is excluded from the competing-phase set, ideal configurational entropy stabilizes only the most dilute composition at 1000 K and 2000 K **(Fig. S12a)**. At absolute equilibrium, the calculation rejects Y, but the margin at the dilute limit is remarkably small (~3 meV/atom at 2000 K, **Fig. S12b**), which falls well within PBE accuracy. This indicates that a dilute fraction of Y can realistically remain trapped as a metastable solid solution. This theoretical prediction of partial Y rejection perfectly aligns with the weak $Y_2O_3$ secondary phase detected in our XRD and the oxidized Y 3d component observed near 158.5 eV in XPS.

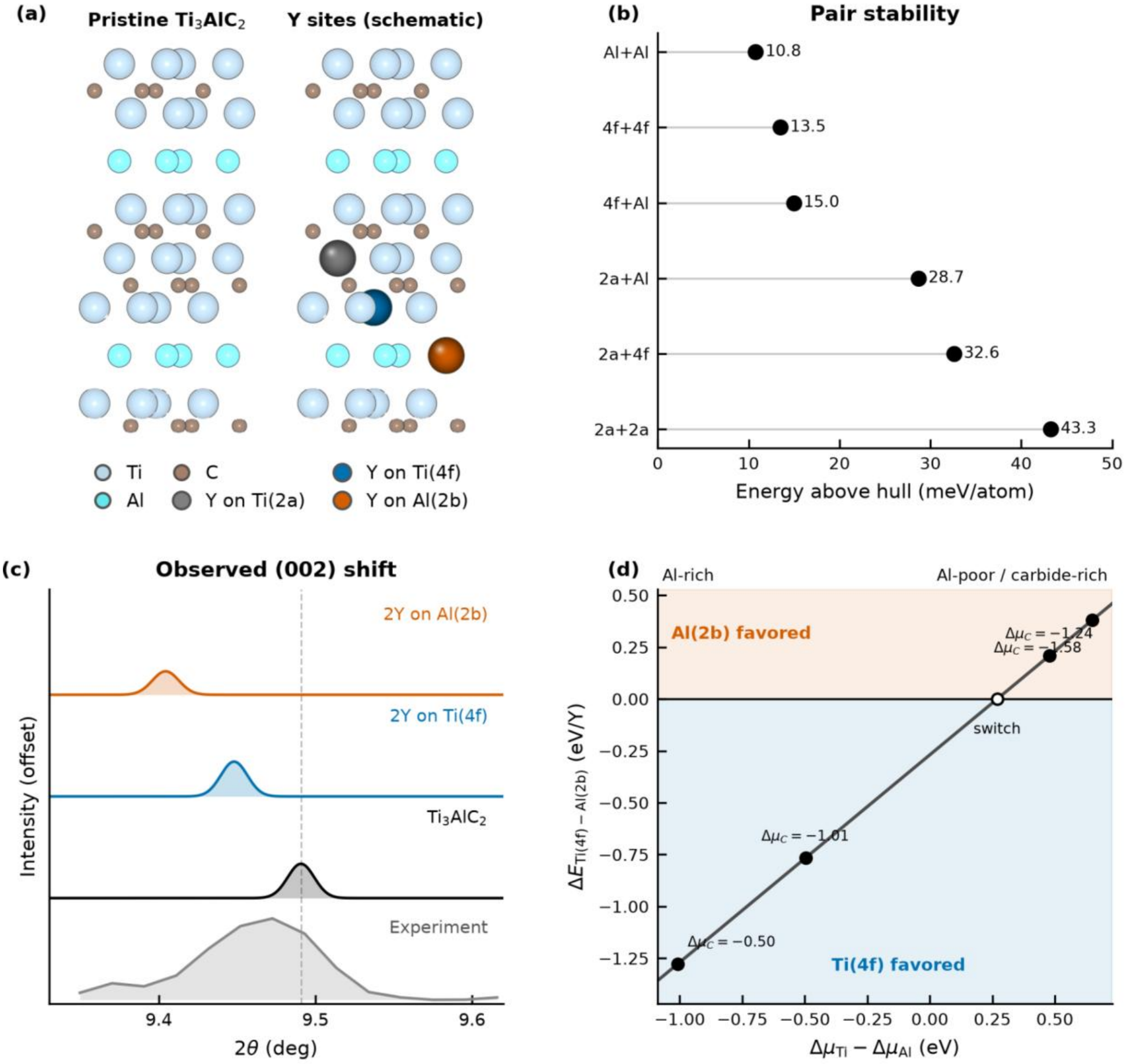

**Figure 4.** DFT results for Y in Ti3AlC2. (a) Pristine $Ti_3AlC_2$ and the three candidate Y sites: gray Ti(2a), blue Ti(4f), and orange Al(2b). (b) Energy above the Ti–Al–Y–C convex hull of the lowest-energy cell of each of the six two-Y site families. (c) Simulated Cu Kα (002) patterns, broadened for display, for pristine $Ti_3AlC_2$ and the two pair models at y = 0.111, together with the measured Y-doped pattern (gray). The experimental trace is shifted so that the undoped (002) position aligns with the DFT pristine peak; only shifts are compared. (d) Ti(4f) minus Al(2b) substitution-energy difference per Y as a function of ΔμTi −ΔμAl over the allowed $Ti_3AlC_2$ domain, following Wang et al. [1]. Negative values favor Ti(4f), positive values favor Al(2b), and the open circle marks equal preference. Black circles are the four Ti–Al–C limiting conditions; labels give the dependent ΔμC at each point.

Conversely, the carbide-like component detected near 156 eV strongly supports the successful incorporation of a smaller Y fraction into a C-coordinated Ti site.

Among the 32 DFT-relaxed two-Y cells, the lowest-energy arrangement in each sublattice family identifies the preferred site (**Table ST4, Fig. 4b**). Within the Ti sublattice ($Y_2Ti_{52}Al_{18}C_{36}$), the Ti(4f)+Ti(4f) pair lies 2.06 eV per cell below the best Ti(2a)+Ti(4f) arrangement and 3.22 eV below Ti(2a)+Ti(2a). In the mixed family ($Y_2Ti_{53}Al_{17}C_{36}$), Ti(4f)+Al(2b) lies 1.48 eV below Ti(2a)+Al(2b). In every family that contains a Ti site, the Ti(4f) member is the minimum. The isolated-Y limit gives the same ranking: one Y on Ti(4f) lies 1.29 eV below one Y on Ti(2a), about nine times $k_BT$ at typical synthesis temperatures, so Ti(2a) occupancy at equilibrium is negligible. The preference is stronger than for Cr, which favors the same Ti(4f) site in $Ti_3AlC_2$ by 0.63 eV [15]. Any Y on the Ti sublattice therefore sits on Ti(4f), the Ti layer facing Al. The one comparison that composition alone cannot settle, Ti(4f) against Al(2b), requires the chemical environment and is treated next. Because the Ti(4f) pair ($Y_2Ti_{52}Al_{18}C_{36}$) and the Al(2b) pair ($Y_2Ti_{54}Al_{16}C_{36}$) have different compositions, their total energies cannot be compared directly. The ranking depends on the chemical potentials of Ti and Al, which vary with synthesis conditions. Following Wang et al. [28], we bounded the allowed chemical potentials of Ti, Al, and C. These potentials reflect the relative abundance of each element during synthesis: an Al-rich powder mixture corresponds to a high Al chemical potential, an Al-poor mixture in contact with TiC to a high Ti chemical potential. $Ti_3AlC_2$ is stable only within a domain where none of the competing phases $TiAl_3$, $Ti_4AlC_3$, $Ti_2AlC$, or $Ti_8C_5$ precipitates. The Ti(4f)–Al(2b) energy difference depends only on $\Delta\mu Ti - \Delta\mu Al$, so this two-dimensional domain reduces to a one-dimension al interval **(Fig. 4d)**. The Y chemical potential cancels because both cells hold two Y. At the Al-rich end, Ti(4f) is favored by 0.77-1.28 eV per Y. At the Al-poor, carbide-saturated end, Al(2b) is favored by 0.21-0.38 eV per Y. Substitution stays endothermic at every allowed condition, with a floor of +0.59 eV per Y. The excess-Al powder mixture used in this work places the synthesis at the Al-rich end. We therefore assign dissolved Y to Ti(4f). The TiC detected in the product marks regions of Al loss, which sit toward the carbide-saturated end where Al(2b) wins by 0.2-0.4 eV per Y. The assignment therefore holds for grains that grew in contact with excess Al, and the (110) scan below tests it directly. This site assignment explains the experimental signatures and lattice response. DFT calculations show that Y on Ti(4f) expands both the a and c lattice parameters ($\triangle a = +0.32\%$, $\triangle c = +0.28\%$ at $y = 0.056$). This matches our XRD data: the larger Y atom on the Ti site pushes apart the basal plane while stretching the interlayer spacing. The Ti(4f) model at $y = 0.056$ predicts a (002) shift of -0.027°, which is in excellent agreement with our measured shift of -0.020° **(Fig. 4c)**, confirming that the incorporated fraction is slightly below the nominal value due to secondary phase formation. The synthesis targets $y = 0.10$ in the starting powder, but $Y_2O_3$ forms as a secondary phase, so the fraction that enters the lattice is smaller. The Ti(4f) model at $y = 0.056$ predicts $\Delta$ (002) = -0.027°; the measured shift is -0.020° (**Fig. 1c**), consistent with an incorporated fraction below the nominal value.

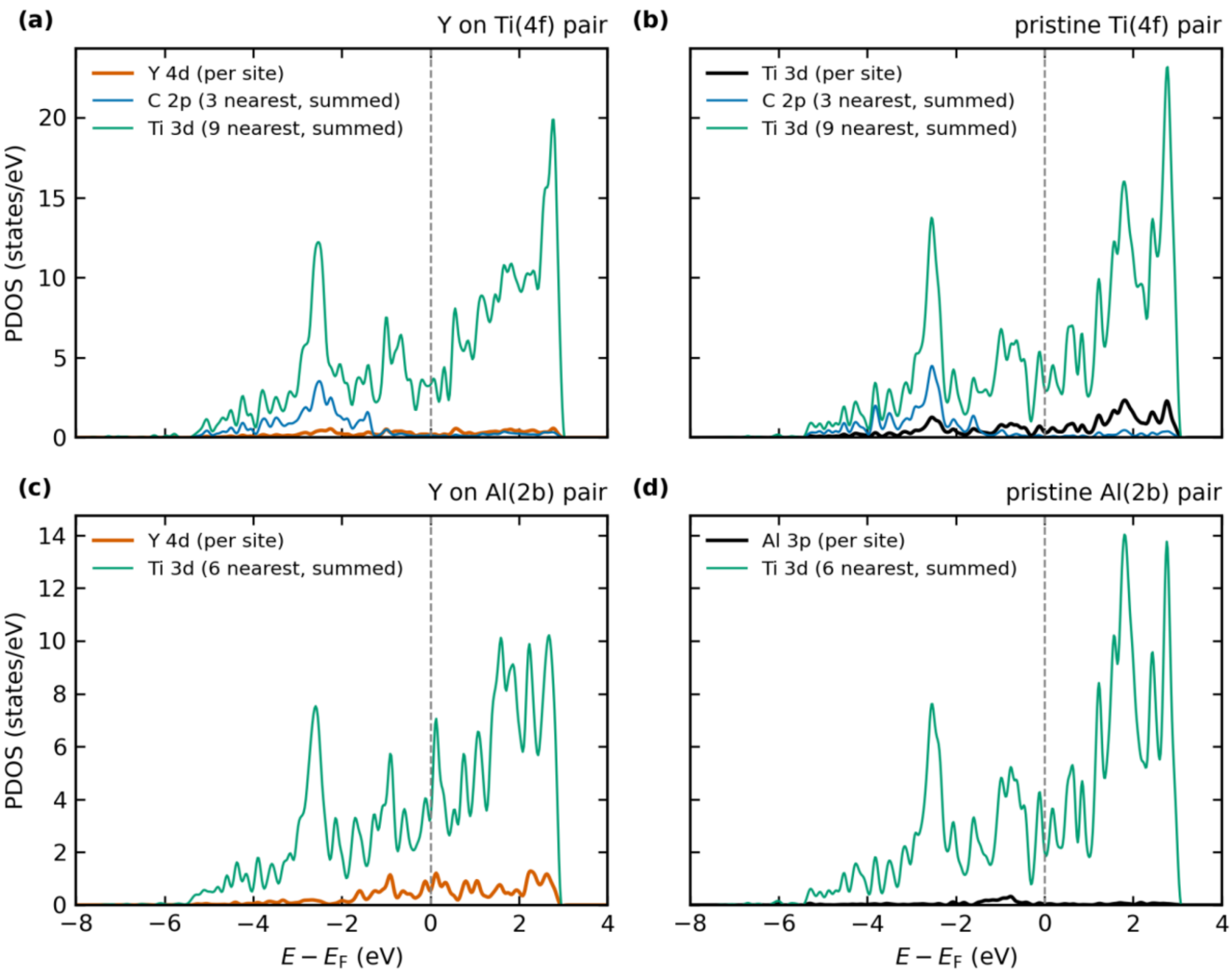

Figure 5. Spin-summed site-projected density of states for the relaxed Ti(4f)-pair and Al(2b)-pair models (left) and the corresponding sites in pristine $Ti_3AlC_2$ (right). Dopant and host curves are averaged per site; neighbor-shell C 2p and Ti 3d curves are summed per central site. Energies are referenced to the Fermi level (dashed line). The absence of a C curve for Y-Al(2b) reflects the lack of carbon within 3.85 ˚ A.

The experimental (104) reflection also shifts toward lower 2θ, in agreement with the lattice expansion predicted by the Ti(4f) model. Because the (00l) lines sense only c, they cannot by themselves confirm the site. The a-sensitive (110) line can: the Ti(4f) model predicts Δ(110) = -0.21° at y = 0.056, while the Al(2b) model predicts only -0.04° (**Table ST5**). The neighboring TiC (220) line stays fixed and serves as an internal reference. A 58-62 step scan on both powders would confirm the Ti(4f) assignment. The carbide-like Y 3d XPS component near 156 eV is consistent with Y on the Ti sublattice, where Y keeps three C neighbors at 2.43 Å on Ti(4f) and six at 2.36 Å on Ti(2a), close to the 2.48 Å Y-C bond in $Y_2C$. On Al(2b) the nearest C sits at 3.85 Å, too far for a carbide-like signal. The oxidized component near 158.5 eV likely reflects surface oxidation or the $Y_2O_3$ secondary phase detected by XRD. The site-projected DOS (**Fig.5**) is consistent with this: Y 4d weight on Ti(4f) spreads over the C 2p region between -5 and -2 eV, whereas on Al(2b) it concentrates near the Fermi level. In summary, DFT calculations confirm that Y exclusively occupies the outer Ti (4f) site under Al-rich synthesis conditions. While thermodynamics predict a slight rejection of Y-consistent with the secondary $Y_2O_3$ phase-the successfully incorporated fraction drives a significant lattice expansion and forms characteristic carbide-like bonds. This site-specific substitution and the resulting framework distortion explain the unusually large interlayer spacing of the MXene.

To demonstrate how this expanded structure enhances lithium-ion transport, we next evaluate the electrochemical performance of the Y-doped $Ti_3C_2T_x$ electrodes. CVA curves at different scan rates exhibit a typical pseudocapacitive shape with gradually increasing current values during scan rate increasing (**Fig.6a**). At relatively low scan rates, charge and discharge peaks are visible and located at around 1.5V and 1.3V, respectively (**Fig.6b**). Taking into account that measured current consists of contributions from diffusion-controlled ion intercalation within the MXene interlayers and surface-controlled processes, including electrical double-layer capacitance and fast surface redox reactions associated with the functionalized MXene surface, the current response can be empirically described by:

$$i(v)=i_c+i_d=av^b$$

or, in logarithmic form

$$log\ i(v)=log\ a+b\ log(v)$$

where $a$ and $b$ are adjustable parameters. The $b$-value is obtained from the slope of the linear dependence of $log\ i(v)$ versus $log(v)$ and provides insight into the dominant charge storage mechanism [1]. The calculated $b$-values are 0.65 for the charge direction and 0.54 for the cathodic peak, indicating that the electrochemical process is mainly controlled by Lithium ion diffusion (**Fig.6c**). This result agrees well with the diffusion contribution calculated from the CV curves based on the methodology described in [2,47]. At a scan rate of 2 mV/s, the diffusion-controlled process contributes about 92% of the total current (**Fig.6d**). As the scan rate increases, the capacitive contribution gradually becomes more important, but diffusion remains the dominant charge storage mechanism, accounting for about 55% of the current even at 200 mV/s. The high diffusion contribution is attributed to the effect of Y doping on the MXene structure and expansion of the interlayer spacing. As discussed in the characterization section, even a small amount of incorporated Yttrium leads to a significant increase of distance between layers of MXene. The expanded interlayer distance provides easier pathways for Li ions, allowing them to diffuse more efficiently.

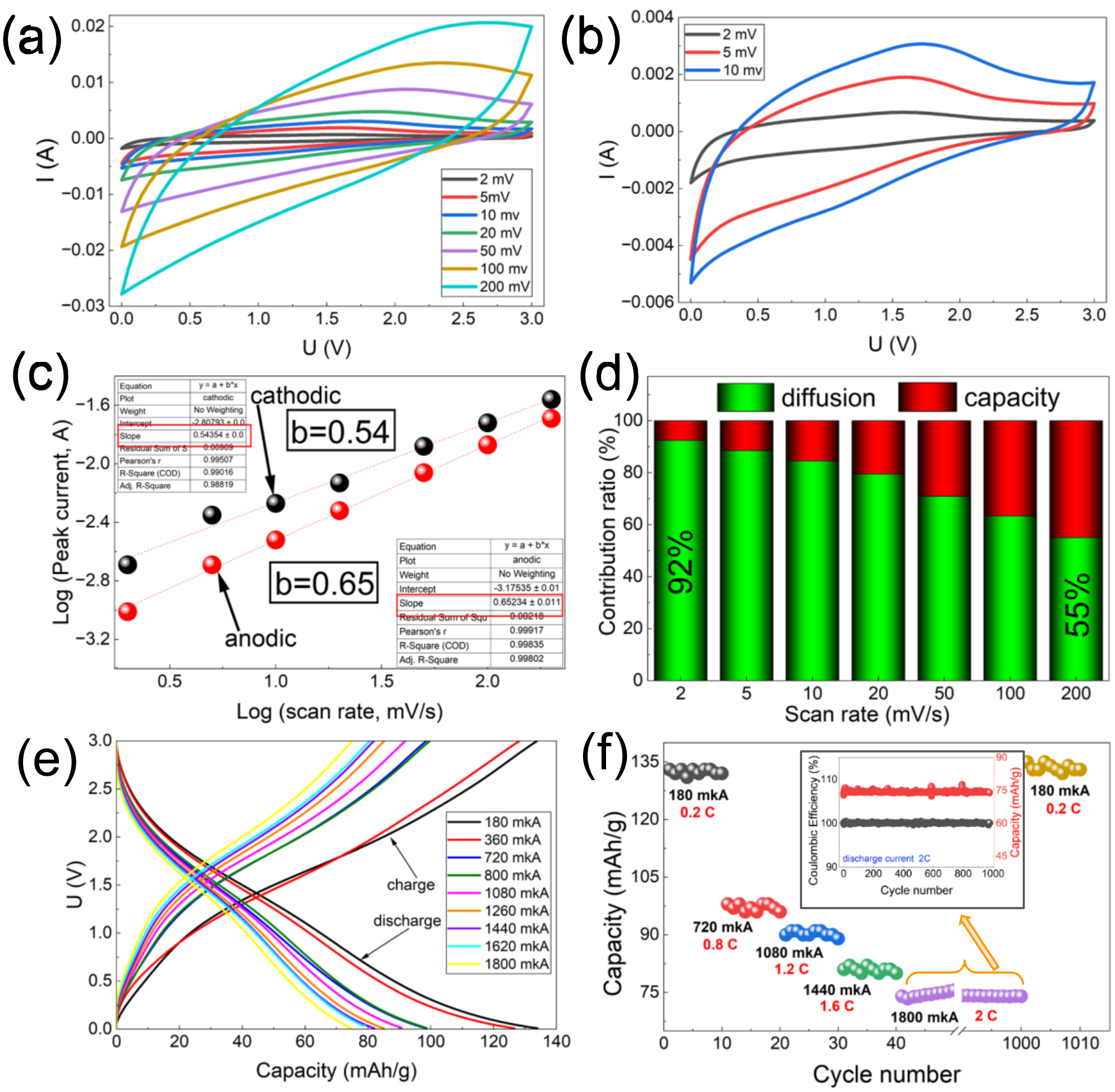

**Figure 6.** Electrochemical performance of Y-doped $Ti_3C_2T_x$ MXene cathode in lithium cells. (a) Cyclic voltammograms recorded at scan rates from 2 to 200 mV/s. (b) Enlarged cyclic voltammogram highlighting the cathodic and anodic peaks. (c) Determination of *b*-values from the relationship between peak current and scan rate. (d) Separation of capacitive- and diffusion-controlled contributions at different scan rates. (e) Galvanostatic charge-discharge curves measured at different current densities. (f) Rate capability and long-term cycling performance of the Y-doped $Ti_3C_2T_x$ cathode.

As a result, the Y-doped MXene exhibits a higher diffusion contribution than our undoped reference sample (about 62% at 5mV/s), as well as multilayered MXenes described in [14] (about 50% at 5mV/s), [48] (about 64% at 0.1 mV/s) or single-flakes-based electrodes in our previous work [7] (56% at 5 mV/s), demonstrating that expanded interlayer distance due to the Y doping promotes ion intercalation into the MXene structure. Charge-discharge curves measured at different current densities (**Fig.6e**) show nearly symmetric charge-discharge profiles with a small horizontal plateau, confirming good reversibility and low internal resistance. The electrode delivers a discharge capacity of approximately 130 mAh/g at 0.2C and retains about 74 mAh/g at 2C (**Fig.6f**). Furthermore, excellent cycling stability is achieved, with a reversible capacity of about 74 mAh/g maintained after 1000 cycles at 2C, while the Coulombic efficiency remains close to 100% throughout the cycling test.

To gain further insight into the ion transport properties of the Y-doped $Ti_3C_2T_x$ electrode, the lithium-ion diffusion coefficient was determined using the galvanostatic intermittent titration technique (GITT) [49-51]. This technique is widely used for the quantitative determination of ion diffusion in electrode materials and provides complementary information to the CV kinetic analysis. GITT measurements were performed at discharge currents of 0.1C (180 mkA), 0.2C (360 mkA), and 0.5C (720 mkA). **Fig. 7a** shows the typical voltage response consisting of a series of discharge pulses followed by relaxation periods. Even at the highest current of 0.4 C, the electrode based on Y-doped MXene exhibits a stable voltage response with a clear relaxation behavior, indicating reversible lithium insertion and extraction. The validity of the GITT analysis was verified by plotting the cell voltage as a function of the square root of time (**Fig.7b**). A nearly perfect linear relationship was obtained for all investigated current densities, demonstrating that the diffusion process satisfies the semi-infinite diffusion assumption required for reliable calculation of the lithium-ion diffusion coefficient [52,53]. This result confirms that the selected current densities are appropriate for GITT analysis.

The quantitative calculation of the Li+ diffusion coefficient was based on the theoretical model of diffusion kinetics and calculated using the following equation [54]

$$D = \frac{4}{\pi\tau} \left(\frac{n_m V_m}{S}\right) \left(\frac{\Delta E_S}{\Delta E_\tau}\right)$$

where $\tau$ is the relaxation time, $n_m$ and $V_m$ represent the molar amount and molar volume of active electrode, respectively; $S$ is the electrode/electrolyte contacting area; $\Delta E_s$ and $\Delta E_\tau$ are the voltage change induced by the pulse operation and during galvanostatic discharge, respectively (**Fig.SI 13**). The calculated lithium-ion diffusion coefficients are presented in **Fig. 7c**. The diffusion coefficient increases gradually during lithiation (discharge) and lies in the range of approximately $10^{-9}$–$10^{-11}$ $cm^2/s$, indicating relatively fast lithium transport within the Y-doped MXene, which is higher than reported for Mxene-based electrodes, as presented in **Table 1**.

Table 1. Calculated values of Li diffusion coefficient compared with undoped Mxene and other reported results

| Electrode | Type of electrode | Working ion | Calculation method | Diffusion coefficient value | Reference |
|---|---|---|---|---|---|
| *Y-doped $Ti_3C_2T_x$* | *cathode* | *Li* | *GITT* | *$10^{-9}$- $10^{-11}$* | *This work* |
| $Ti_3C_2$ film | cathode | Mg | GITT | $10^{-11}$- $10^{-14}$ | [55] |
| 3D interwoven $Ti_3C_2$ MXene networks | cathode | Mg | GITT | $4.27 \times 10^{-12}$ | [55] |
| Interlayer-expanded $Ti_3C_2T_x$ | cathode | Li | GITT | $2.85\times 10^{-13}$ | [56] |
| $Ti_2C$ MXene on carbon cloth | anode | Li | GITT | $10^{-11}$- $10^{-12}$ | [57] |
| Combustion-synthesized $Ti_3AlC_2$ | anode | Li | GITT | $6.6 \times 10^{-10}$ | [58] |
| N, P co-doped $Ti_3C_2T_x$ | anode | Li | GITT | $10^{-9}$- $10^{-12}$ | [59] |

As expected, the apparent diffusion coefficient decreases with increasing discharge current because of higher polarization. Nevertheless, even at a current rate of C/2, the diffusion coefficient remains within the same order of magnitude, indicating that expanded interlayer spacing induced by Y incorporation provides efficient pathways for lithium-ion transport. Furthermore, the nearly constant diffusion coefficient between 2.0 and 1.0 V (horizontal plateau in **Fig.7c**) suggests stable diffusion kinetics throughout the main lithiation process. The GITT results are further supported by the EIS measurements (**Fig.7d**). The Nyquist plots recorded at different cell voltages exhibit small semicircles in the high-frequency region, followed by inclined diffusion tails at low frequencies. The impedance spectra were analyzed by fitting the experimental data with an equivalent circuit model (**Fig.SI 14**), consisting of elements representing the SEI resistance, charge-transfer resistance, and lithium-ion diffusion (Warburg). Both $R_{SEI}$ and $R_{ct}$ gradually decrease during lithiation, indicating the formation of a stable electrode/electrolyte interface and faster charge-transfer kinetics as lithium ions are inserted into the cathode material [60]. At the same time, the Warburg coefficient decreases from approximately 75 to 25 $\Omega/s^2$, indicating lower diffusion resistance and easier lithium-ion transport (**Fig.7e**). These results are in good agreement with the GITT analysis, where the lithium-ion diffusion coefficient remains in the range of $10^{-9}$–$10^{-11}$ $cm^2/s$ throughout the discharge process.

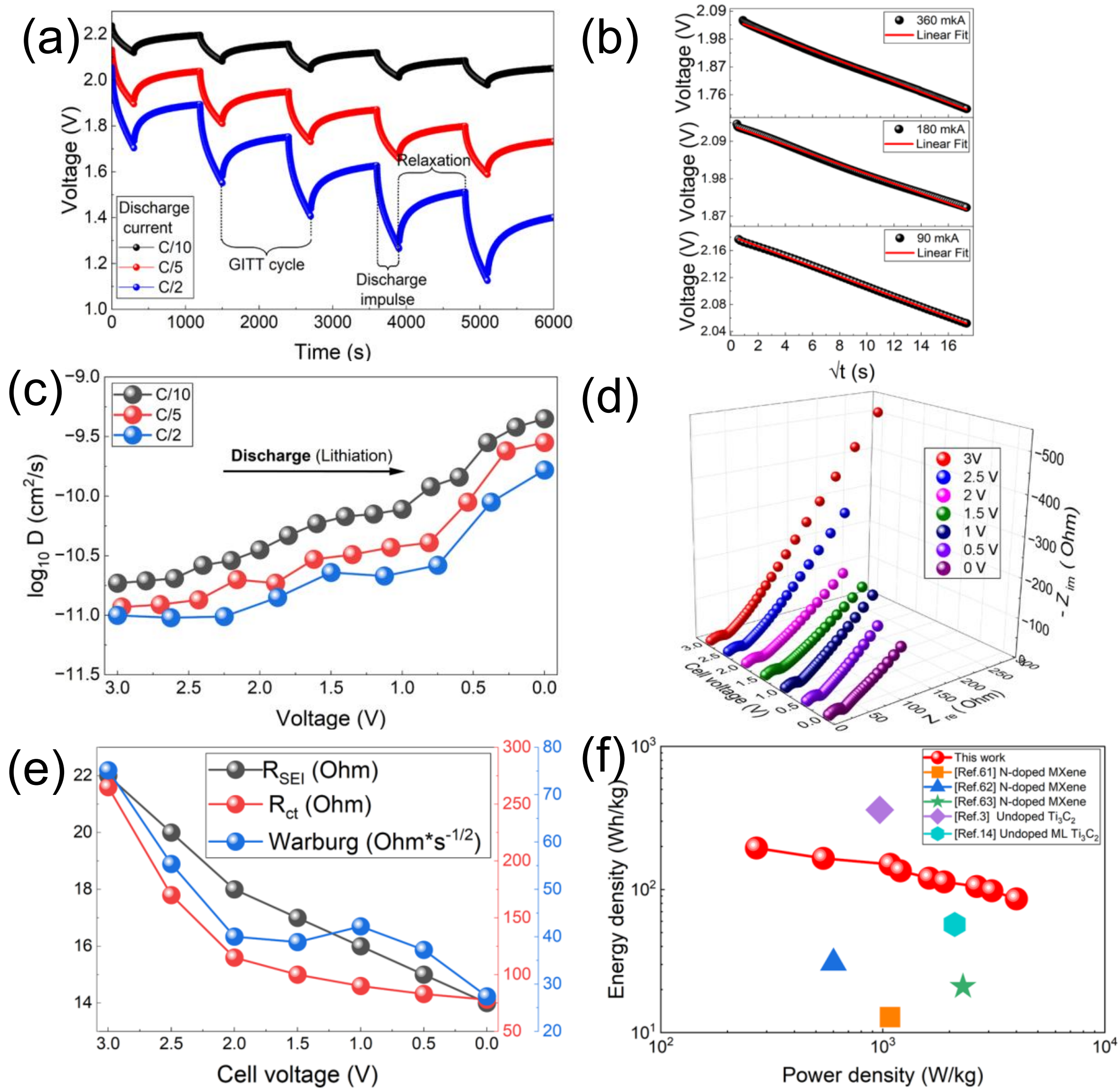

**Figure 7.** Lithium-ion diffusion kinetics and impedance analysis of Y-doped $Ti_3C_2T_x$ MXene. (a) Representative GITT voltage profile recorded during discharge at different current rates. (b) Linear relationship between cell voltage and the square root of time, confirming semi-infinite diffusion conditions. (c) Lithium-ion diffusion coefficient calculated from GITT as a function of cell voltage. (d) Nyquist plots measured at different discharge voltages. (e) Evolution of fitted charge-transfer resistance ($R_{ct}$), SEI resistance ($R_{SEI}$), and Warburg coefficient during discharge. (f) Ragone plot, showing energy-power performance compared to published result

Moreover, the nearly constant diffusion coefficient between 2.0 and 1.0 V suggests stable lithium-ion transport during discharge. The improved diffusion behavior can be directly related to the structural changes introduced by Y incorporation. As confirmed by XRD, Raman, XPS, and SEM-EDS analyses, Y modifies the $Ti_3C_2T_x$ structure, increases the interlayer spacing, and promotes the formation of hydroxyl-rich surface terminations. The electrochemical metrics summarized in **Table ST6** confirm that Y incorporation effectively enhances the rate capability of $Ti_3C_2T_x$ MXene compared to our undoped sample, resulting in superior capacity retention and higher energy and power densities at high current rates. Furthermore, the Ragone plot (**Fig. 7f**) demonstrates that Y-doped $Ti_3C_2T_x$ MXene achieves a higher power density than previously reported MXene-based electrodes. These results indicate that substitutional doping of the MAX precursor is an effective strategy for tailoring the interlayer spacing and ion-transport properties of MXenes, providing a promising route toward the rational design of high-power energy-storage electrodes.

## Conclusions

In this work, we demonstrated a new approach for engineering $Ti_3C_2T_x$ MXene through substitutional Y doping of the parent $Ti_3C_2T_x$ MAX phase. Unlike reported conventional post-synthesis modification methods of MAX phase during etching, Y incorporation into the MAX lattice induces structural changes that are retained after etching, leading to a significant expansion of the MXene interlayer spacing. DFT calculations confirm that Y preferentially occupies the outer Ti(4f) site facing the Al layer. While thermodynamic models predict partial Y rejection as $Y_2O_3$, the successfully incorporated fraction drives a local lattice distortion that facilitates MAX phase etching and promotes hydroxyl surface functionalization. Comprehensive structural and kinetic characterization confirms that this targeted interlayer expansion, combined with the hydroxyl-rich surface chemistry, dramatically accelerates lithium-ion transport within the resulting Y-doped Mxene. Y-doped $Ti_3C_2T_x$ cathode exhibits diffusion-controlled charge storage, a high $Li^+$ diffusion coefficient ($10^{-9}$-$10^{-11}$ $cm^2/s$), excellent cycling stability (74 mAh/g after 1000 cycles at 2C), and enhanced power density (up to 3970 W/kg) compared with undoped MXene and previously reported multilayered MXene cathodes. These findings establish MAX-phase substitutional doping as an effective strategy for controlling the interlayer structure and ion-transport kinetics of MXenes, opening new opportunities for the rational design of high-power electrodes for next-generation energy storage systems.

## Acknowledgments

The present work has been supported by the projects MFA2022/009 (SPINO2D), Prometeo/2021/082, and CIPROM/2024/4 of the Generalitat Valenciana, the TED2021-132656B-C22 granted by the Call 2021 - "Ecological Transition and Digital Transition Projects" promoted by the Ministry of Science and Innovation, funded by the European Union within the "NextGeneration" EU program, the Recovery, Transformation, and Resilience Plan, and the State Investigation Agency, and the funding from Agencia Estatal de Investigación through the Project PID2023-146181OB-I00 (UTOPIA). M. E. C. acknowledges financial support from the Santiago Grisolia Program of the Generalitat Valencia (grant CIGRIS/2021/150). A.B and T.B acknowledge financial support from the University of Valencia (through the Program "Universitat de València with Ukraine", the Generalitat Valenciana through the Program "Acoge CV-UCRANIA personal investigador" of the Conselleria de Innovación, Universidades, Ciencia y Sociedad Digital, and the Ministry of Universidades through the program "Plan de Acción Universidad-Refugio".

**Declaration of Interests**

The authors declare that they have no known competing financial interests or personal relationships that could have appeared to influence the work reported in this paper

**Data availability**

Data will be made available on request.

# Supplementary information (SI)

## Site-Selective Yttrium Substitution in $Ti_3AlC_2$ Enables Interlayer Engineering and $Li^+$ Transport in $Ti_3C_2T_x$ cathodes for High-Power Energy Storage

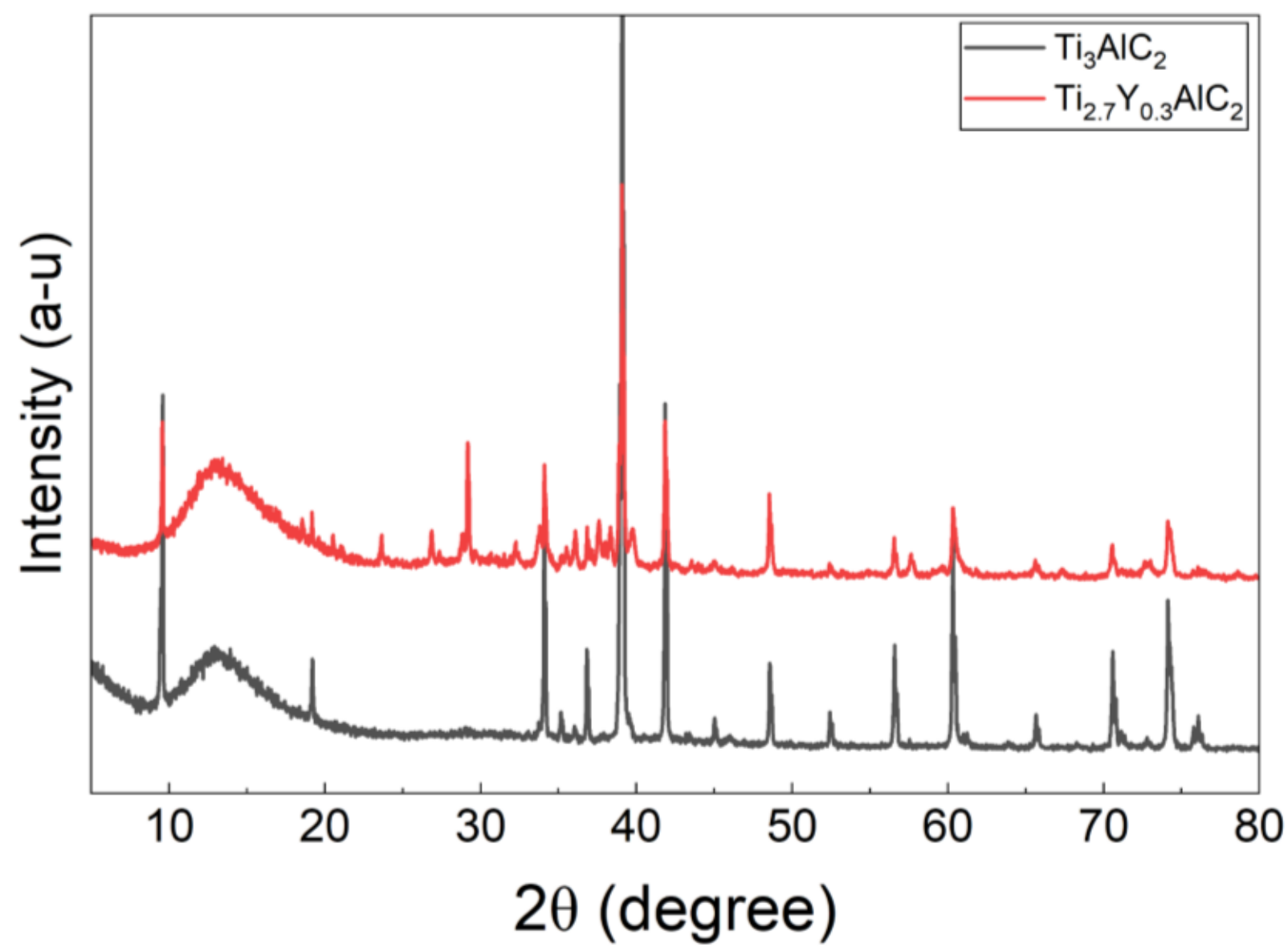

**Figure SI 1.** XRD pattern of undoped $Ti_3AlC_2$ MAX phase used as the reference sample.

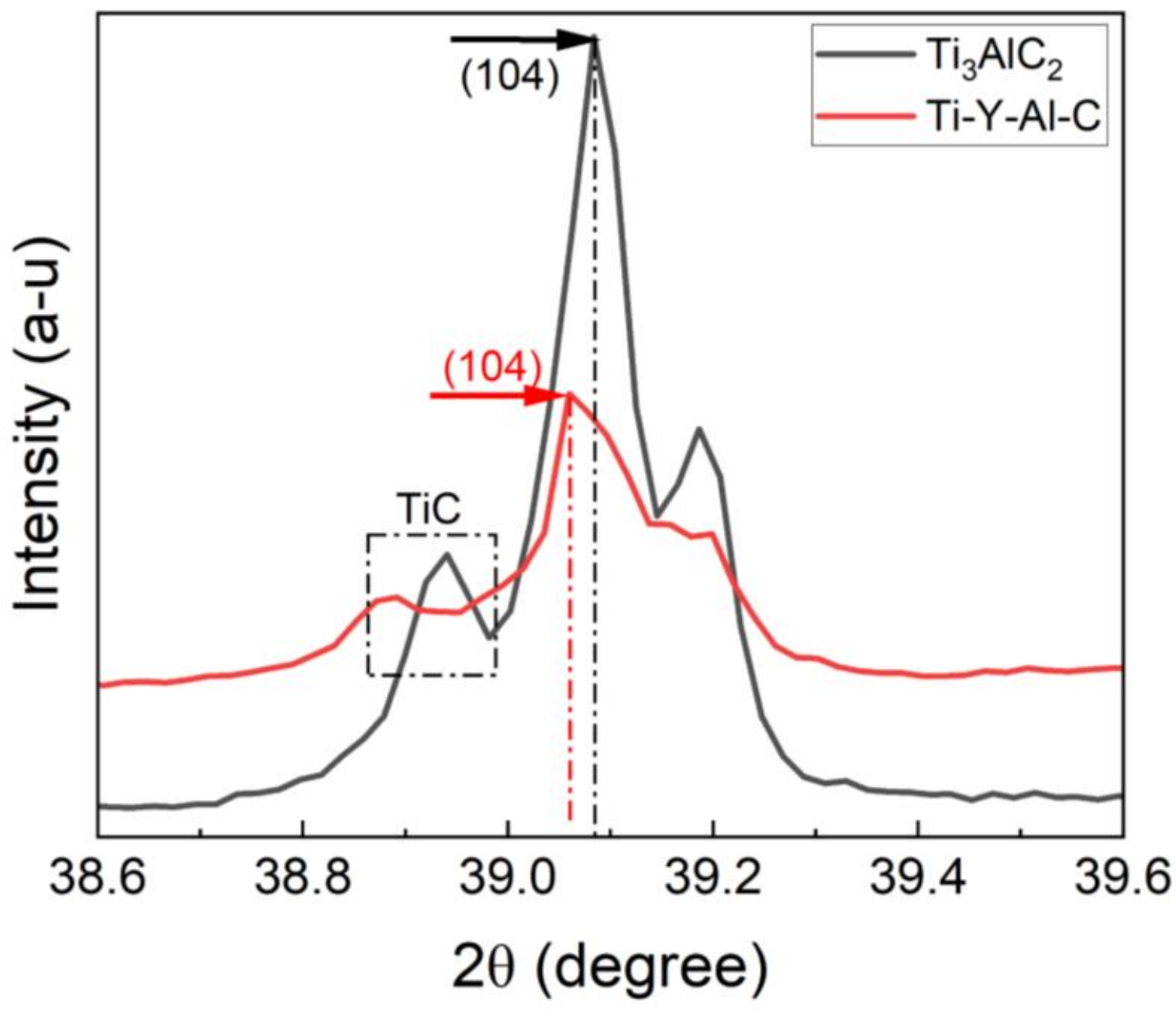

**Figure SI 2.** Enlarged XRD patterns in the region of the (104) reflection for pristine $Ti_3AlC_2$ and Y-doped (Ti–Y–Al–C) MAX phases.

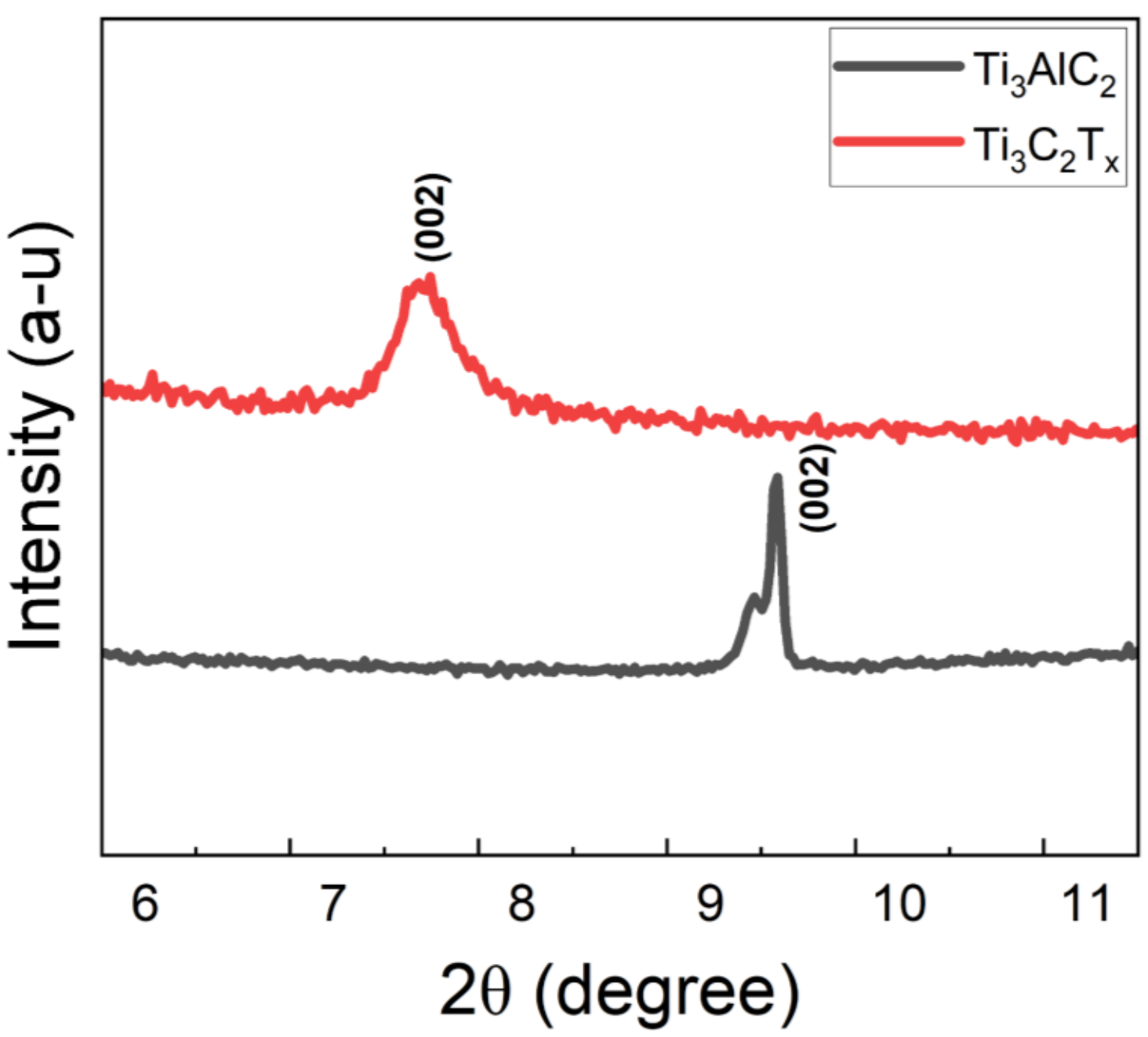

**Figure SI 3.** Comparison of the (002) diffraction peak shift after etching for undoped $Ti_3C_2T_x$ MXene, illustrating the increase in interlayer spacing relative to the parent MAX phase.

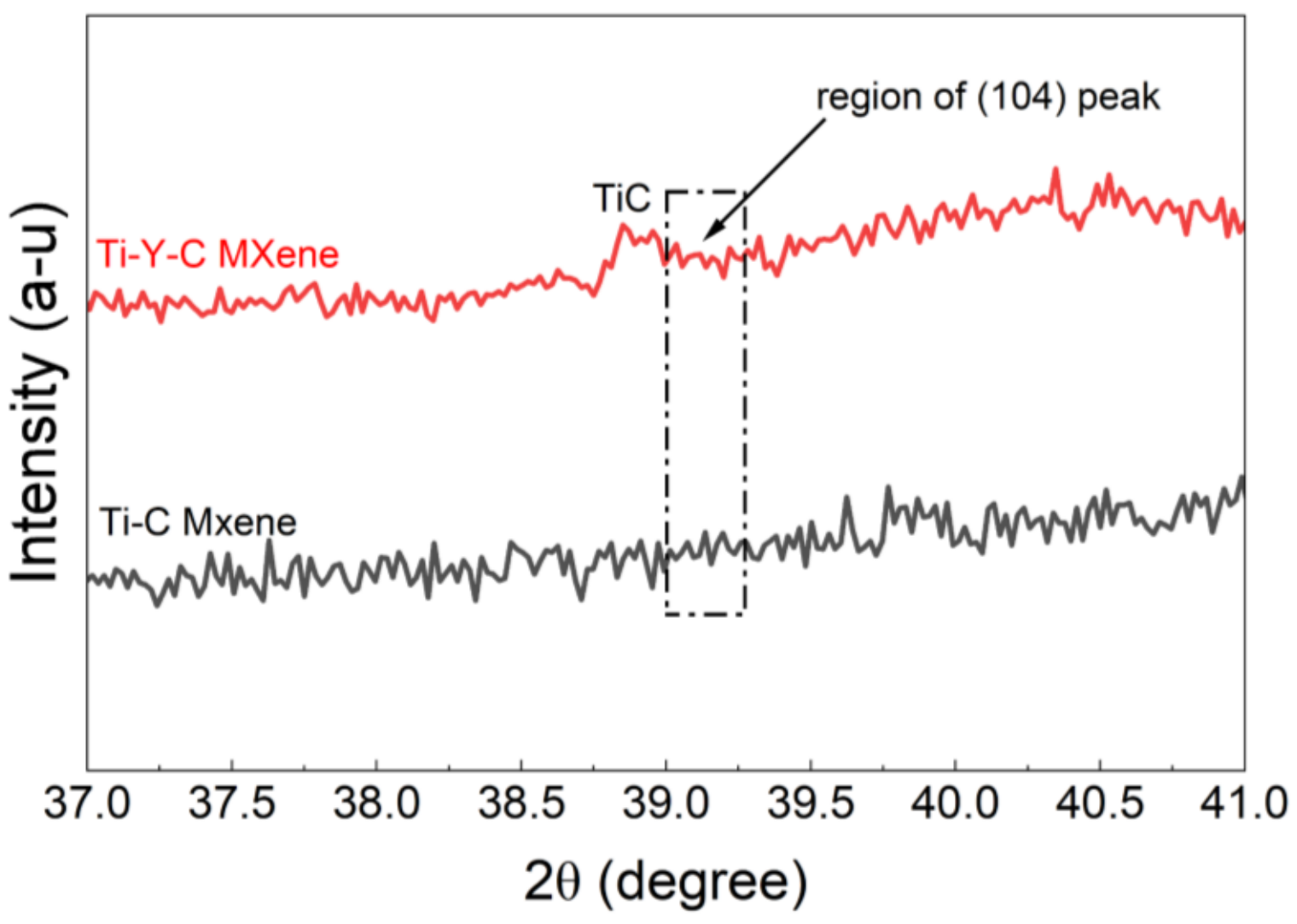

**Figure SI 4.** Enlarged XRD patterns of etched undoped $Ti3C_2T_x$ and Y-doped MXene in the 37–41° 2θ range. The absence of the characteristic MAX (104) reflection in both samples confirms the successful transformation of MAX into MXene after selective Al removal. The residual TiC reflection is indicated.

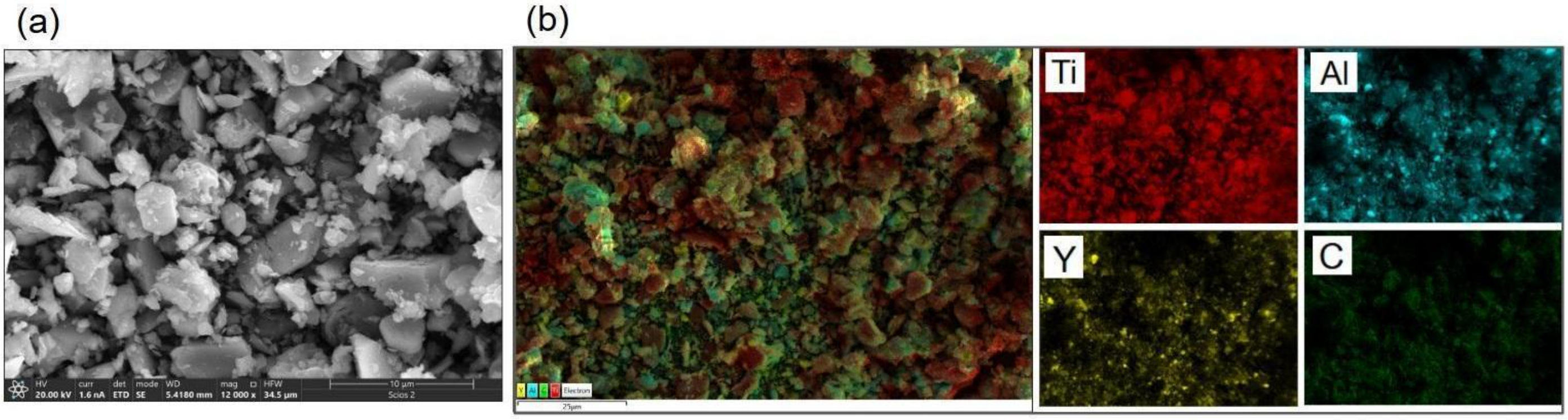

**Figure SI 5.** Additional SEM characterization of Y-doped $Ti_3AlC_2$ MAX phase: (a) particle size distribution and (b) EDS elemental mapping demonstrating homogeneous elemental distribution.

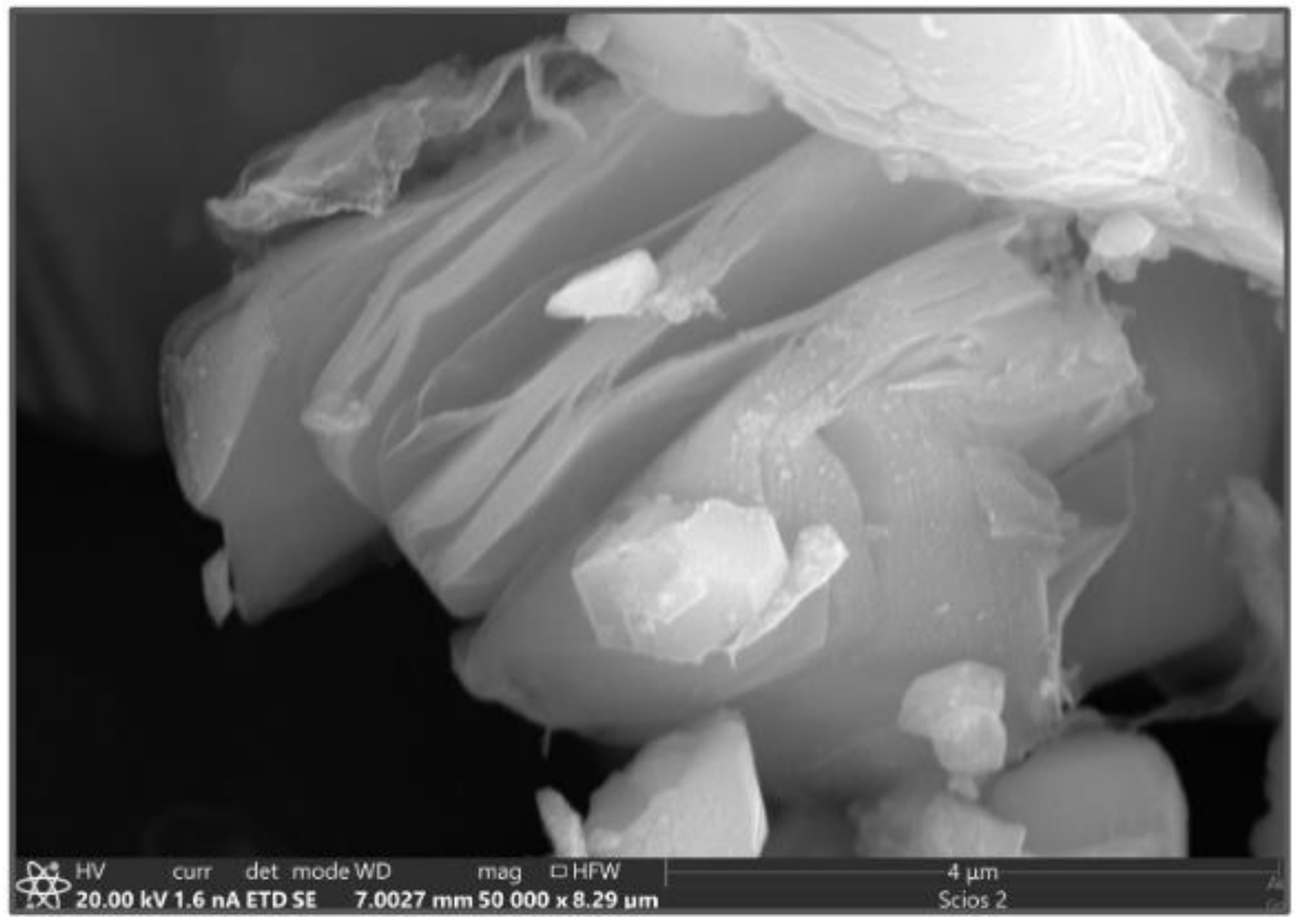

**Figure SI 6.** SEM images of multilayered Y-doped $Ti_3C_2T_x$ MXene showing the preserved lateral flake dimensions after chemical etching.

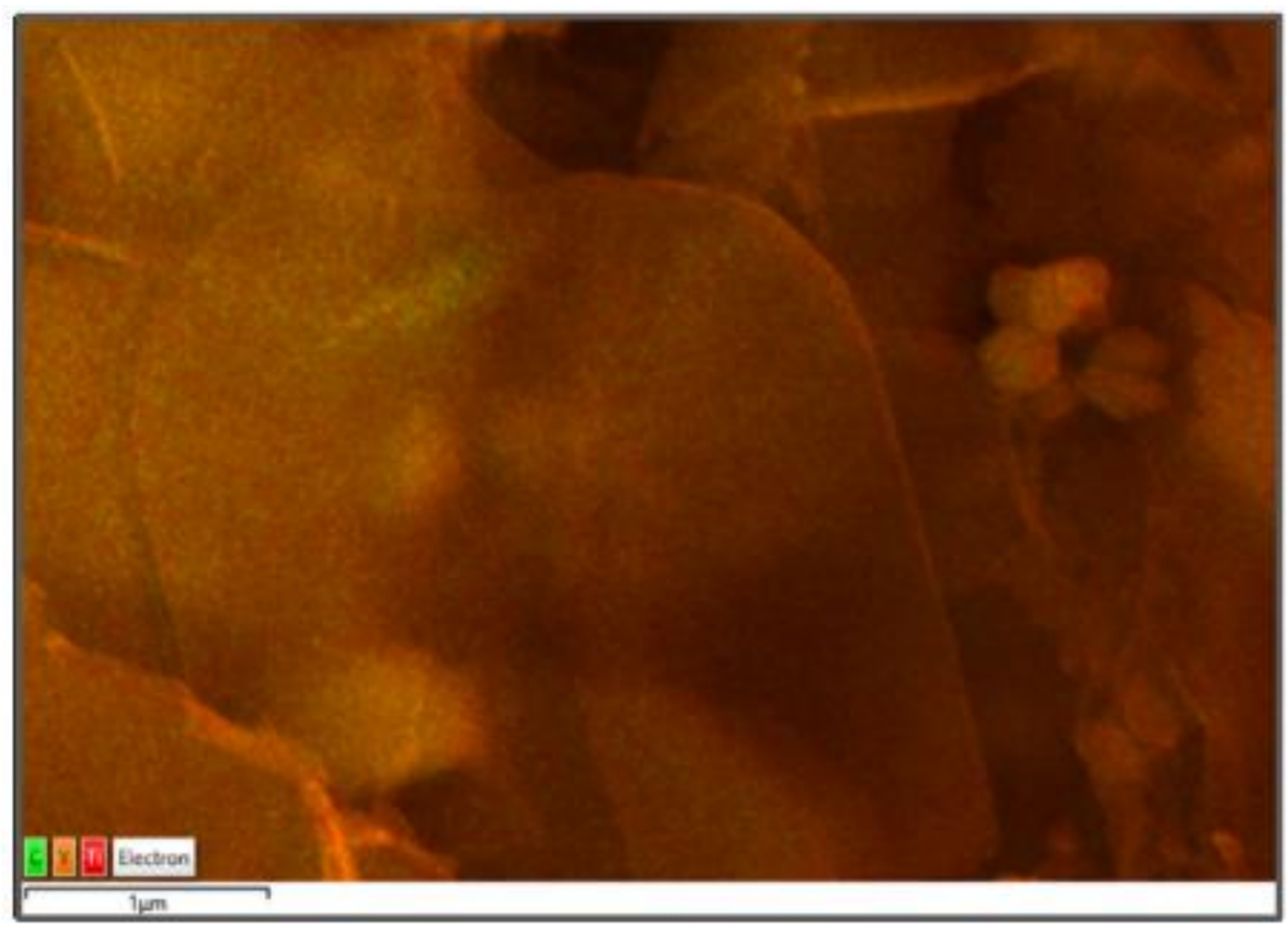

**Figure SI 7.** EDS elemental mapping of an individual delaminated Y-doped $Ti_3C_2T_x$ MXene flake, confirming homogeneous distribution of Y throughout the flake.

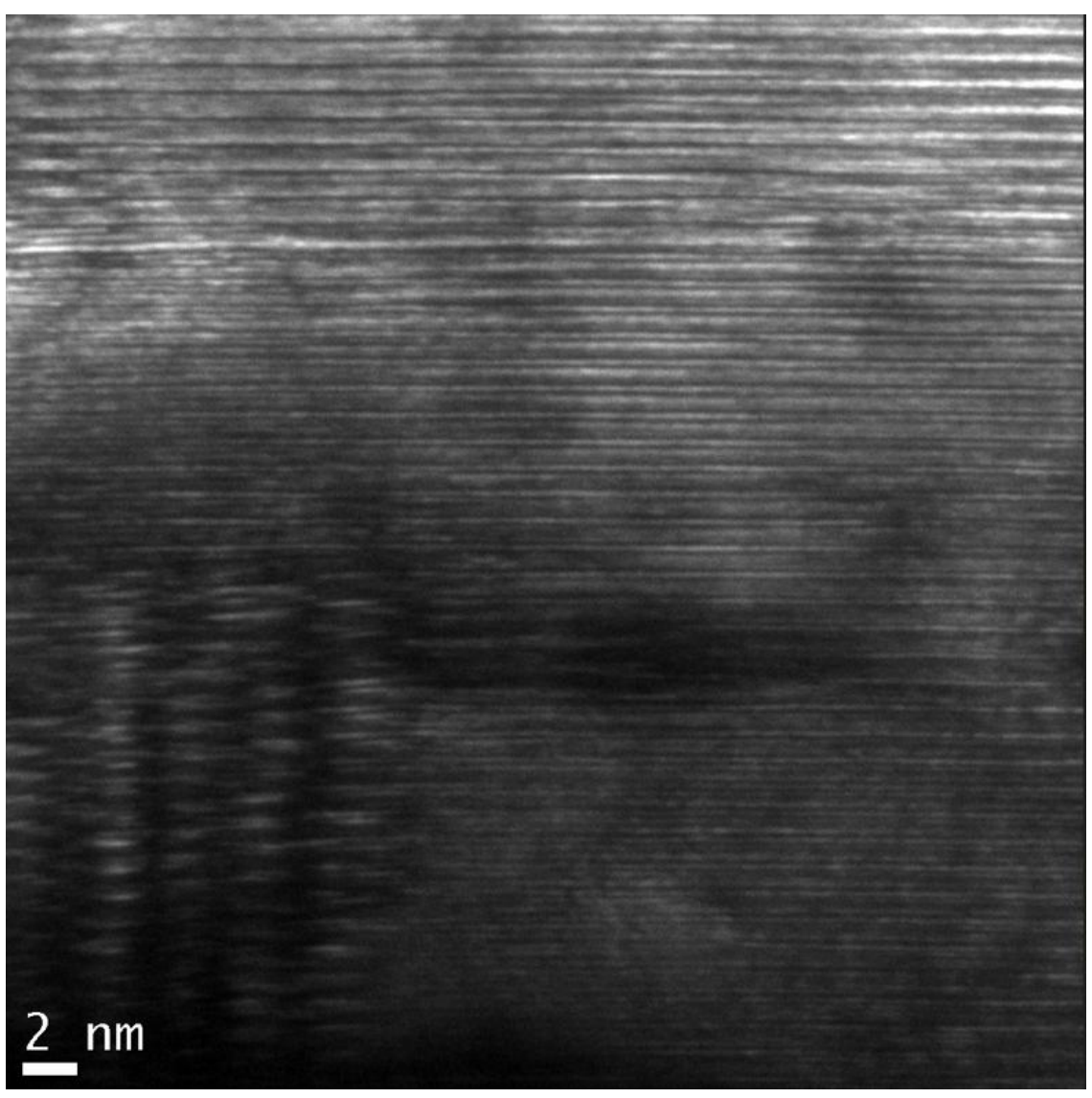

**Figure SI 8.** High-resolution TEM image of the Y-doped $Ti_3AlC_2$ MAX phase showing the layered crystal structure and measured interlayer spacing.

**Table SI 1.** Quantitative EDX elemental composition of Y-doped $Ti_3AlC_2$ MAX phase and Y-doped $Ti_3C_2T_x$ MXene before and after etching.

| MAX | | | |
|---|---|---|---|
| Element | Weight % | Atomic % | Ti:Y |
| AlK | 15.7 | 25.4 | 28.8 |
| TiK | 79.1 | 72.0 | |
| Y K | 5.2 | 2.5 | |
| Total | 100 | 100 | |

| MXene | | | |
|---|---|---|---|
| Element | Weight % | Atomic % | Ti:Y |
| TiK | 93.8 | 96.6 | 28.4 |
| Y K | 6.2 | 3.4 | |
| Total | 100 | 100 | |

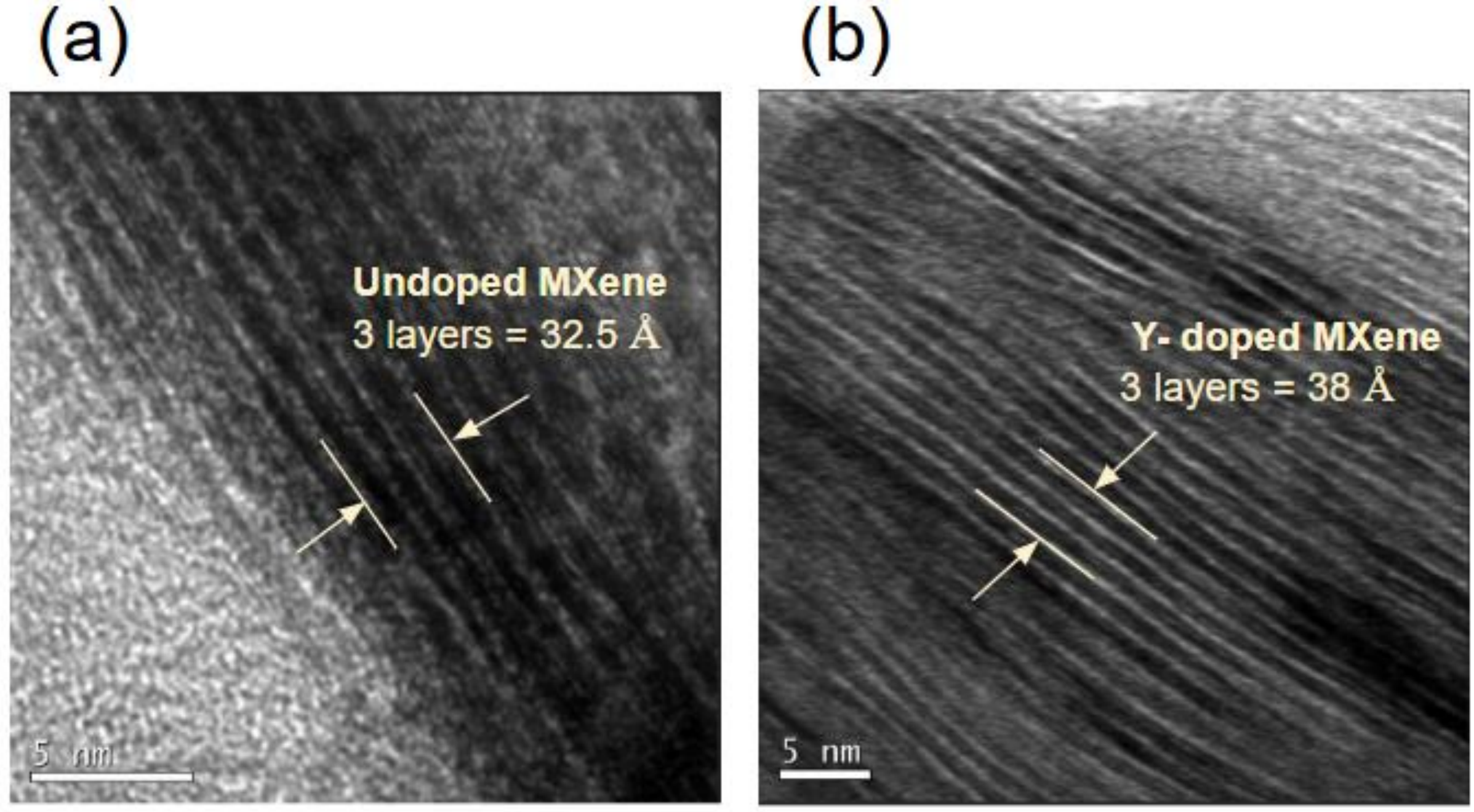

**Figure SI 9.** Comparison of interlayer spacing between undoped and Y-doped multilayered $Ti_3C_2T_x$ MXenes obtained from TEM analysis.

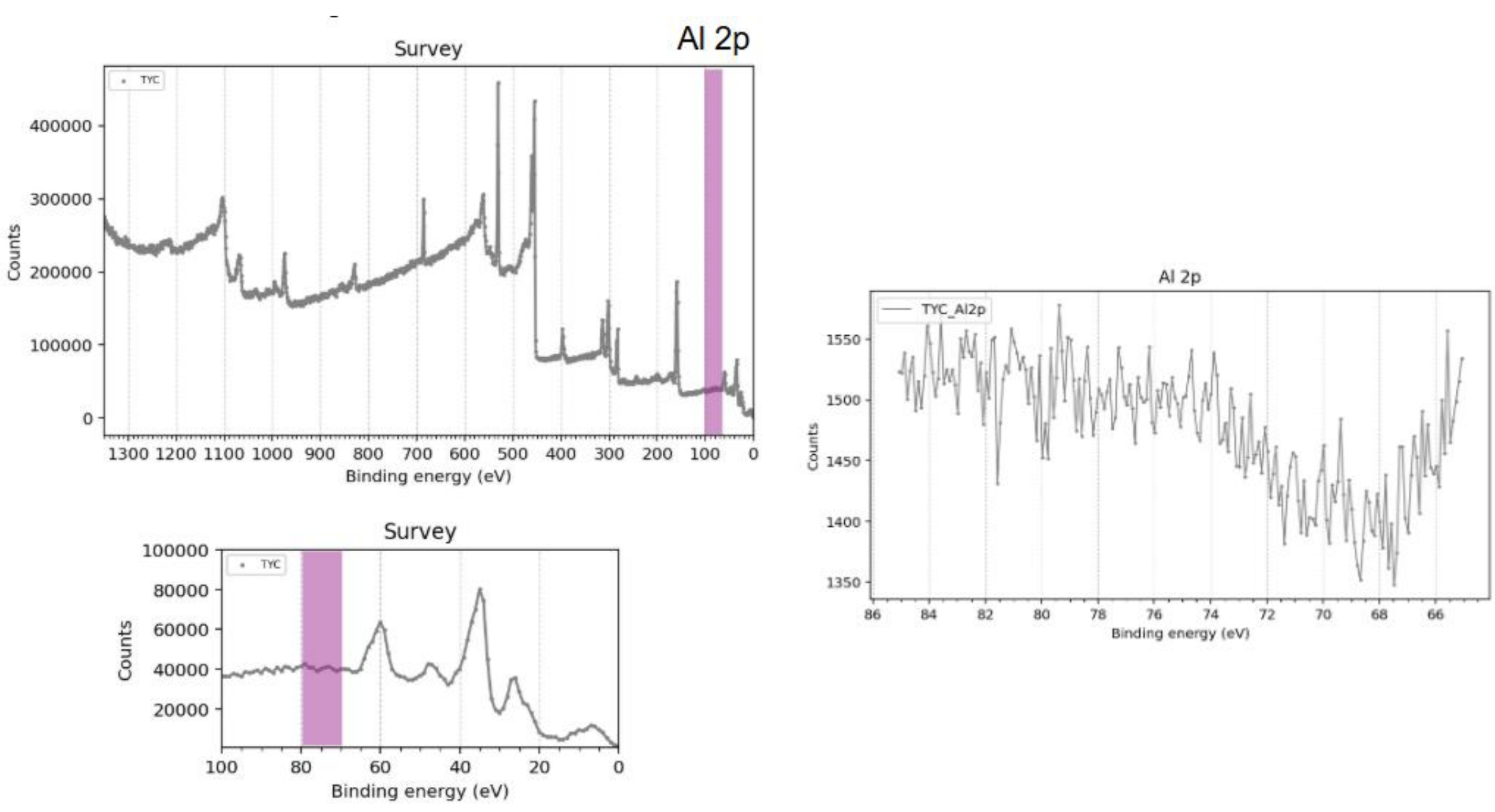

**Figure. SI 10.** XPS survey spectra of Y-doped $Ti_3C_2T_x$ MXene after etching, confirming the elemental composition and the absence of detectable Al.

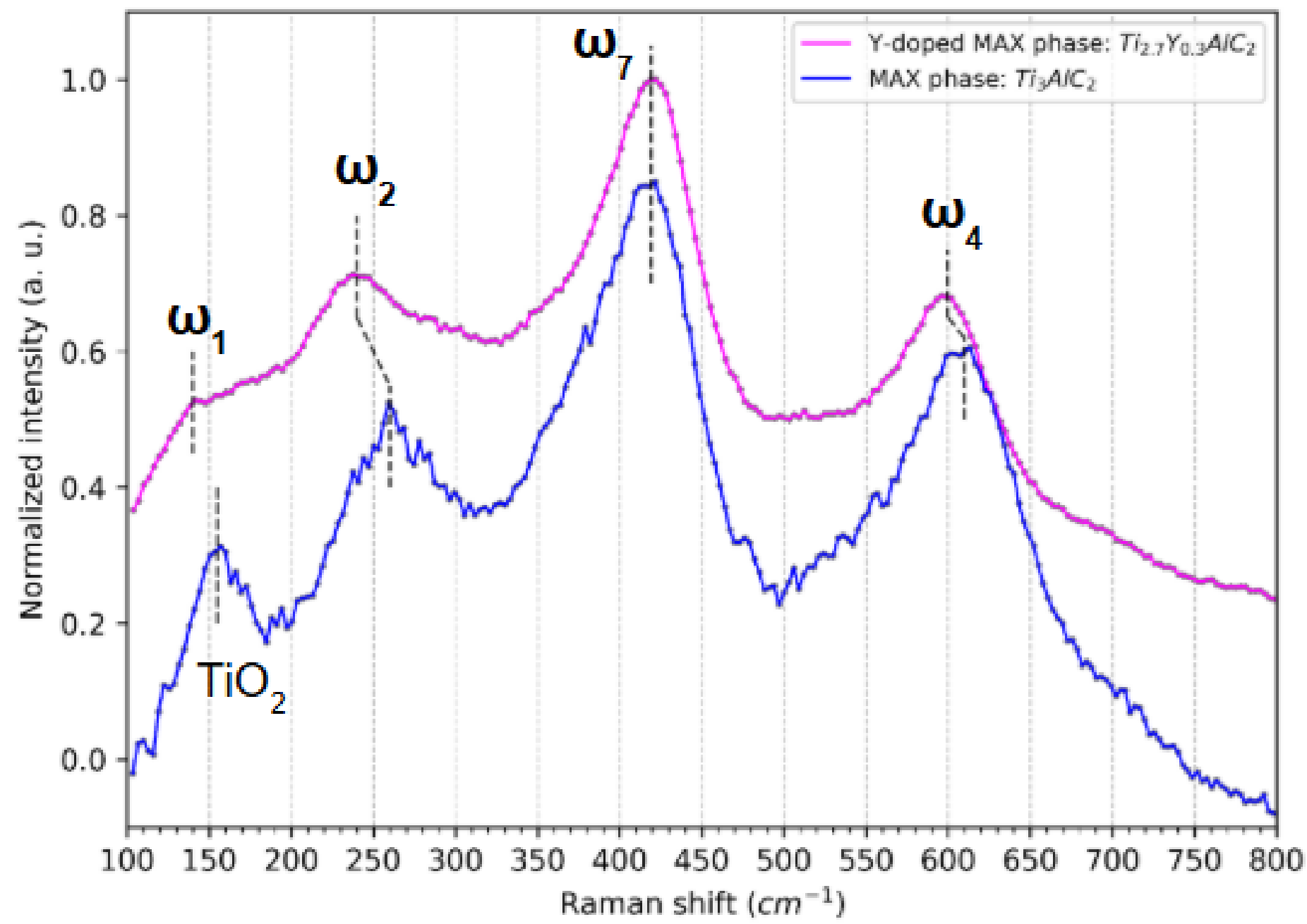

**Figure. SI 11.** Comparison of Raman spectra of Y-doped $Ti_3AlC_2$ MAX phase and Y-doped $Ti_3C_2T_x$ MXene, highlighting changes in characteristic vibrational modes after etching.

Table ST2. Substitution energies per Y (eV/Y) with elemental reservoirs (hcp Ti, fcc Al, hcp Y), compared with the Cr solution energies of Burr et al.

| Model | $y$ | Site(s) | $E_{\mathrm{sub}}$ (eV/Y) | Burr's Cr (eV) |
|---|---|---|---|---|
| Single Y | 0.056 | Ti(4f) | +1.646 | +1.28 |
| Single Y | 0.056 | Ti(2a) | +2.934 | +1.91 |
| Single Y | 0.056 | Al(2b) | +1.445 | – |
| Two Y | 0.111 | Ti(4f)+Ti(4f) | +1.173 | – |
| Two Y | 0.111 | Ti(4f)+Al(2b) | +1.426 | – |
| Two Y | 0.111 | Al(2b)+Al(2b) | +1.435 | – |

Table ST3. Competing phases used to construct theTi–Al–Y–C convex hull. All energies were computed with PBE, PAW projectors, and a 520 eV cut off at the same settings as the supercells

| Phase | $N$ | $E/N$ (eV) | Phase | $N$ | $E/N$ (eV) |
|---|---|---|---|---|---|
| Al | 1 | −3.753 | Y | 2 | −6.432 |
| C | 2 | −9.224 | Ti | 3 | −7.807 |
| TiC | 2 | −9.327 | $Ti_2C$ | 12 | −8.925 |
| $Ti_8C_5$ | 13 | −9.078 | $Al_4C_3$ | 7 | −6.188 |
| TiAl | 2 | −6.179 | $TiAl_2$ | 12 | −5.529 |
| $TiAl_3$ | 4 | −5.158 | $Ti_3Al$ | 8 | −7.069 |
| $Ti_2AlC$ | 8 | −7.849 | $Ti_3AlC$ | 5 | −7.870 |
| $Ti_3AlC_2$ | 12 | −8.366 | $Ti_4AlC_3$ | 16 | −8.609 |
| $Ti_5Al_2C_3$ | 10 | −8.151 | $Y_2C$ | 3 | −7.676 |
| $Y_3C_4$ | 70 | −8.370 | $Y_4C_5$ | 18 | −8.332 |
| $Y_4C_7$ | 44 | −8.515 | YAl | 4 | −5.513 |
| $YAl_2$ | 6 | −5.174 | $YAl_3$ | 8 | −4.857 |
| $Y_2Al$ | 12 | −5.856 | $Y(AlC)_3$ | 14 | −6.779 |
| $Y_3AlC$ | 5 | −6.892 | $Y_3AlC_3$ | 14 | −7.652 |
| $Y_2TiAl_3$ | 6 | −5.767 | $Y_6Ti_4Al_{43}$ | 106 | −4.702 |

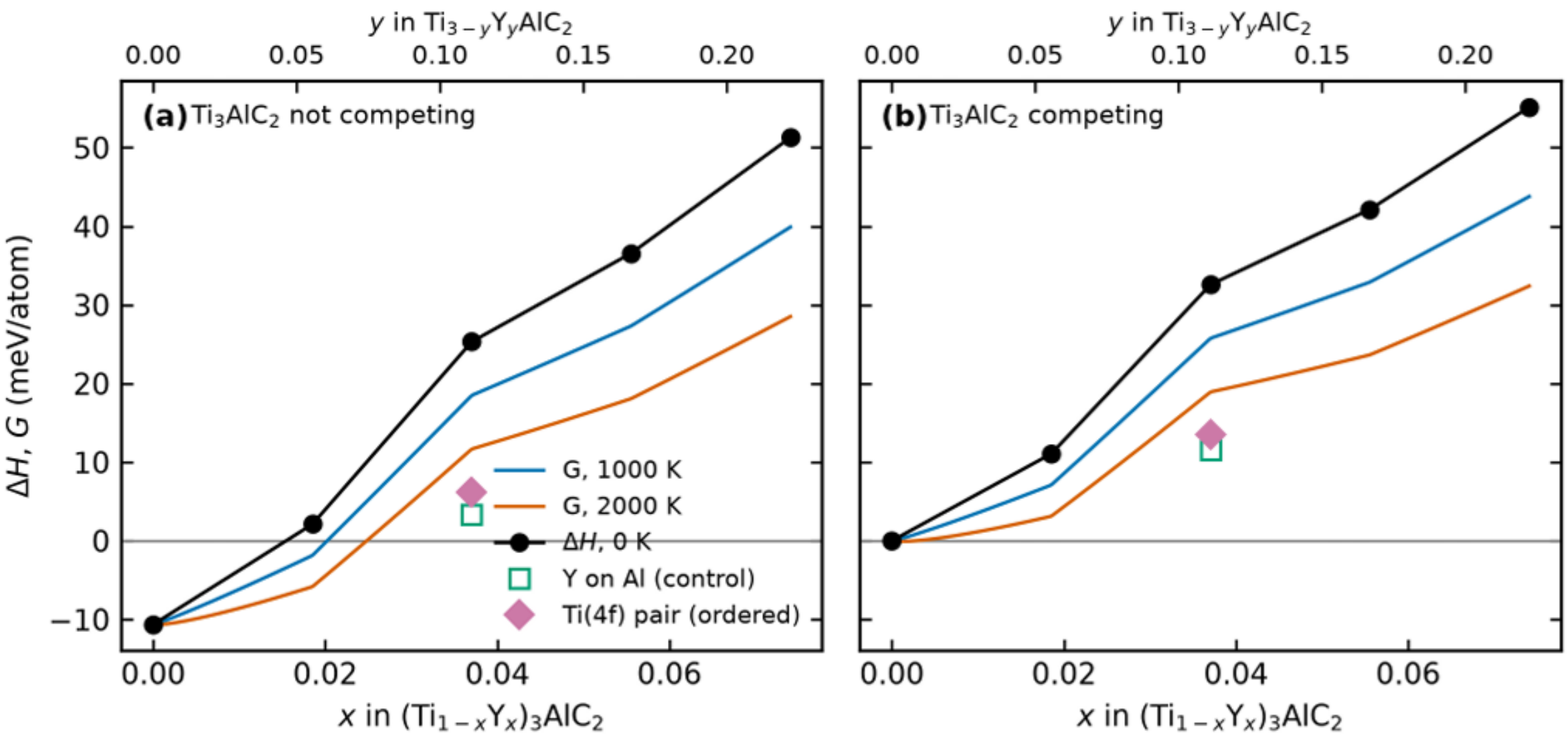

Fig. SI 12. Concentration-dependent stability of the Ti-sublattice SQS series, following Azina et al. [5 C. Azina et al., J. Am. Ceram. Soc. 106 (2023), doi:10.1111/jace.18931]. (a) Formation enthalpy ΔH at 0 K and estimated free energy G = ΔH −TΔSconf at 1000 and 2000 K when Ti3AlC2 is excluded from the competing-phase set. (b) The same with Ti3AlC2 included. Here x is the fraction of Ti sites occupied by Y and y = 3x. The open square is an Al-sublattice control at the same total Y content as y = 0.111; the filled diamond is the ordered Ti(4f) pair at y = 0.111, which avoids the Ti (2a) penalty present in the SQS series.

Table ST4. DFT minimum of each two-Y pair family. ΔE is the family minimum above the lowest structure of the same composition, per Y; a value of 0 marks the winner of its composition group.

| Pair family | $\Delta E$ (meV/Y) | DFT cells |
|---|---:|---:|
| Ti(2a)+Ti(2a) | 1608.1 | 3 |
| Ti(2a)+Ti(4f) | 1031.9 | 3 |
| Ti(4f)+Ti(4f) (Ti-pair model) | 0.0 | 10 |
| Ti(2a)+Al(2b) | 741.6 | 4 |
| Ti(4f)+Al(2b) | 0.0 | 6 |
| Al(2b)+Al(2b) (Al-pair model) | 0.0 | 6 |

Table ST5. Relaxed lattice parameters of the 108-atom cells (parent-cell a, c), their change relative to pristine $Ti_3AlC_2$, and calculated Cu Kα peak shifts (degrees 2θ) relative to pristine $Ti_3AlC_2$ for all DFT cells. All values are direct DFT results from the relaxed cells at the stated y. PBE overestimates c, so only shifts are compared with experiment.

| Model | $y$ | $a$ (Å) | $c$ (Å) | $\Delta a$ (%) | $\Delta c$ (%) | Δ(002) | Δ(008) | Δ(104) | Δ[(104) − (008)] | Δ(110) |
|---|---|---|---|---|---|---|---|---|---|---|
| $Ti_3AlC_2$ | 0 | 3.0812 | 18.6375 | – | – | – | – | – | – | – |
| 1Y on Ti(4f) | 0.056 | 3.0911 | 18.6902 | +0.32 | +0.28 | −0.0269 | −0.1134 | −0.1259 | −0.0125 | −0.21 |
| 1Y on Ti(2a) | 0.056 | 3.0913 | 18.7009 | +0.33 | +0.34 | −0.0323 | −0.1363 | −0.1334 | +0.0029 | −0.22 |
| 1Y on Al(2b) | 0.056 | 3.0832 | 18.7430 | +0.07 | +0.57 | −0.0536 | −0.2263 | −0.0757 | +0.1506 | −0.04 |
| 2Y on Ti(4f) | 0.111 | 3.1004 | 18.7214 | +0.62 | +0.45 | −0.0427 | −0.1801 | −0.2338 | −0.0537 | −0.42 |
| 2Y on Al(2b) | 0.111 | 3.0873 | 18.8082 | +0.20 | +0.92 | −0.0864 | −0.3645 | −0.1501 | +0.2144 | −0.13 |

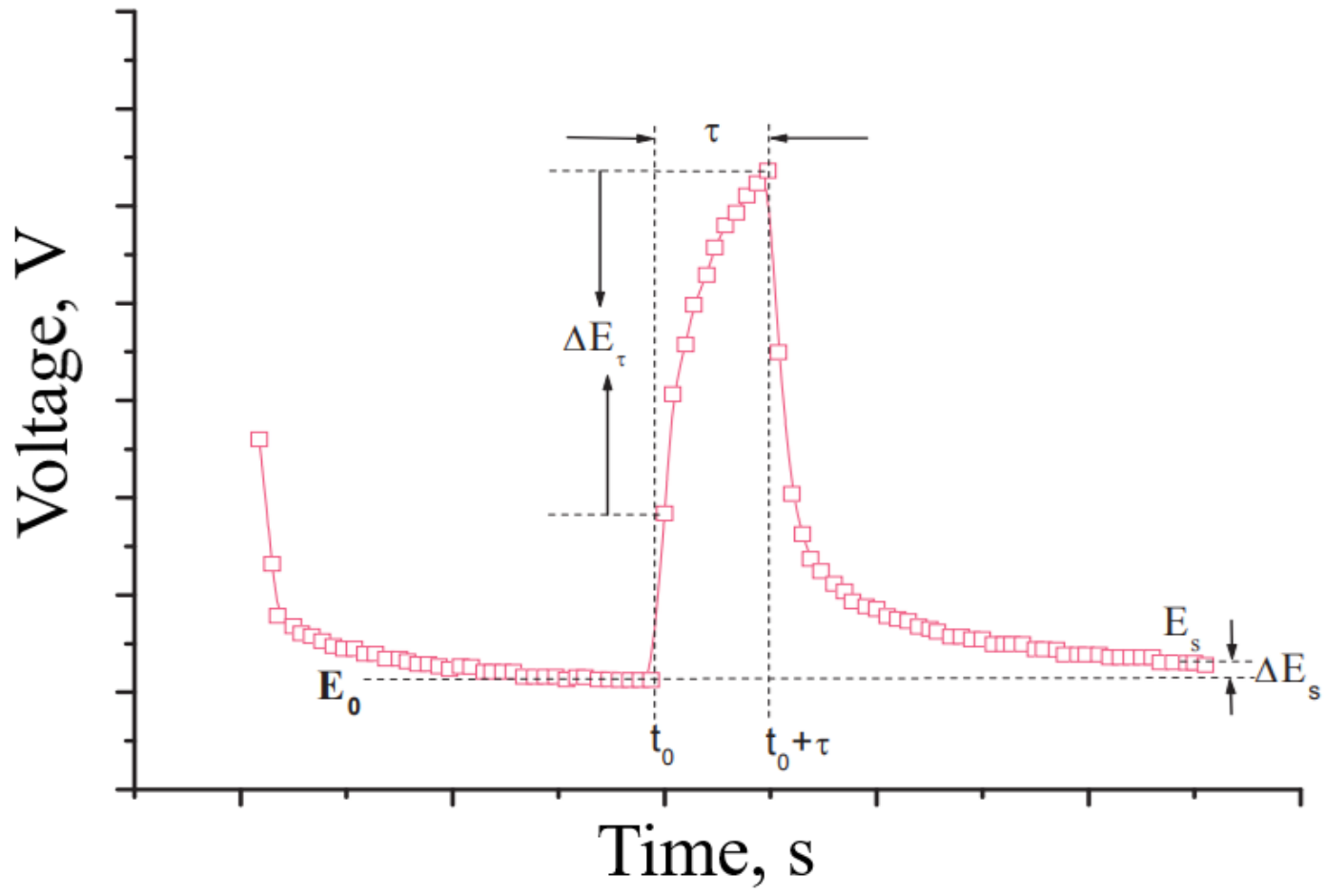

**Figure. SI 13.** Schematic illustration of the parameters used for calculating the lithium-ion diffusion coefficient from GITT measurements, showing the definitions of ΔEτ and ΔEs.

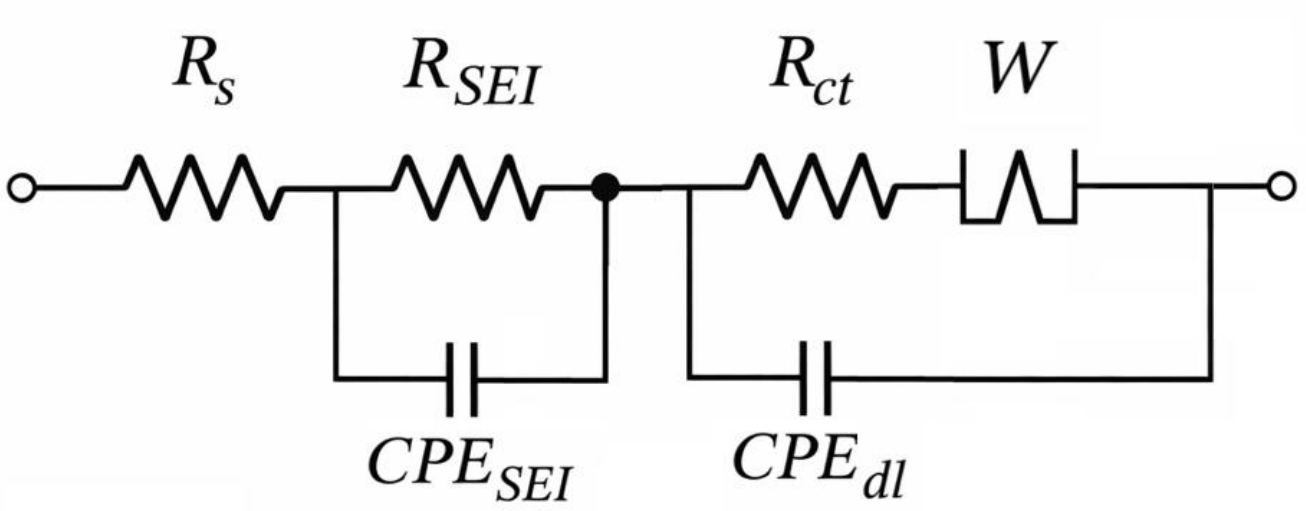

**Figure. SI 14.** An equivalent electrical circuit is used to fit the electrochemical impedance spectroscopy (EIS) data collected during discharge.

**Table ST 6.** Comparison of electrochemical performance, including specific capacity, energy density, and power density, of undoped and Y-doped $Ti_3C_2T_x$ MXene cathodes within a 0–3 V voltage window.

| | Voltage window | MXene loading in the | Current rate | Capacity, Ah/kg | Energy density, Wh/kg | Power density W/kg |
|---|---|---|---|---|---|---|

| | | **electrode** | | | | |
|---|---|---|---|---|---|---|
| **Undoped Mxene** | 0-3V | 4.1 mg/cm2 | 0.2C | 132 | 199 | 220 |
| | | | 2C | 70 | 83 | 3140 |
| **Y-doped MXene** | 0-3V | 3.9 mg/cm2 | 0.2C | 134 | 195 | 270 |
| | | | 2C | 74 | 86 | 3970 |